\documentclass[letterpaper,10 pt, conference]{ieeeconf}  
\IEEEoverridecommandlockouts                              

\usepackage{xcolor}
\usepackage{amsmath} 
\usepackage{amssymb}
\usepackage{amsfonts}

\usepackage{amsthm}
\usepackage{graphicx,url} 
\usepackage{bm}
\usepackage{float}
\usepackage{makecell}
\let\labelindent\relax
\usepackage{enumitem}
\usepackage{subcaption}
\usepackage{wrapfig}
\usepackage{algorithm,algorithmicx} 
\usepackage{algpseudocode} 
\usepackage{multirow}
\usepackage{diagbox}

\usepackage{cite}
\usepackage{soul}  
\usepackage[bookmarks=true]{hyperref}
\hypersetup{
    colorlinks=true,
    pdfborderstyle={/S/U/W 1},
    linkcolor=blue,
    filecolor=magenta,      
    urlcolor= blue,
    linkbordercolor=red,
}
\newcommand\rurl[1]{%
  \href{http://#1}{\nolinkurl{#1}}%
}

\usepackage{comment}

\let\abs\relax
\newcommand{\abs}[1]{\left\lvert#1\right\rvert}
\newcommand{\norm}[1]{\left\lVert#1\right\rVert}

\newcommand{\infnorm}[1]{\left\lVert#1\right\rVert_\infty}
\newcommand{\linfnorm}[1]{\left\lVert#1\right\rVert_{\mathcal{L}_\infty}}

\newcommand{\linfnormtruc}[2]{\left\lVert#1\right\rVert_{\mathcal{L}_\infty^{[0,#2]}}}

\newcommand{\lonenorm}[1]{\left\lVert#1\right\rVert_{\mathcal{L}_1}}

\def\lone{{\mathcal{L}_1}}

\def\laplace#1{\mathfrak{L}\left[#1\right]}

\def \loneAC {$\lone$AC}

\def\nt {\textup{n}}

\def\tilx{\tilde{x}}

\def\dotx{\dot x}

\def \hsigma{\hat{\sigma}}
\def\xin{x_\textup{in}}

\def\xr{x_\textup{r}}

\def \rt {\textup{r}}
\def \rhoin {\rho_\textup{in}}
\def \Zn {\mathbb{Z}_1^n}
\def \Zm {\mathbb{Z}_1^m}

\def \xn {x_\textup{n}}

\def \un {u_\textup{n}}

\def\mbR{\mathbb{R}}

\def\mbZ{\mathbb{Z}}

\def\mbZ{\mathbb{Z}}

\def\hsigma{\hat{\sigma}}

\def\mcC{\mathcal{C}}

\def\mcX{\mathcal{X}}
\def\mcH{\mathcal{H}}
\def\mcU{\mathcal{U}}
\def\mcG{\mathcal{G}}

\def\trieq{\triangleq}

\newtheorem{lemma}{Lemma}

\theoremstyle{definition}  
\theoremstyle{definition} 
\newtheorem{assumption}{Assumption}
\theoremstyle{remark}  
\newtheorem{remark}{Remark}

\usepackage{accents}

\makeatletter
\def\cl@part {\@elt {chapter}}
\makeatother

\usepackage{cleveref}

\crefname{equation}{}{} 
\crefname{lemma}{Lemma}{Lemmas}
\crefname{theorem}{Theorem}{Theorems}
\crefname{table}{Table}{Tables}
\crefname{figure}{Fig.}{Figs.}
\crefname{remark}{Remark}{Remarks}
\crefname{assumption}{Assumption}{Assumptions}
\crefname{section}{Section}{Sections}
\crefname{definition}{Definition}{Definitions}
\crefname{algorithm}{Algorithm}{Algorithms}
\crefname{proposition}{Proposition}{Propositions}

\makeatletter
\renewcommand*\env@matrix[1][\arraystretch]{%
  \edef\arraystretch{#1}%
  \hskip -\arraycolsep
  \let\@ifnextchar\new@ifnextchar
  \array{*\c@MaxMatrixCols c}}
\makeatother

\def\mbZ{\mathbb{Z}}

\def\mcC{{\mathcal{C}}}
\def\mcY{{\mathcal{Y}}}
\def\mcZ{{\mathcal{Z}}}

\def \mcU{{\mathcal{U}}}

\def \loneAC {$\lone$AC}

\usepackage{hyperref}
\usepackage{xcolor}
\hypersetup{
    colorlinks=true,
    pdfborderstyle={/S/U/W 1},
    linkcolor= blue,
    filecolor=magenta,      
    urlcolor= cyan,
    linkbordercolor=red,
    citecolor  = blue
}

\def \xcheckin  {{\check x}_\textup{in}}
\newcommand{\Tau}{\mathrm{T}}
\def \xcheckn {{\check x}_\textup{n}}

\def \at {\textup{a}}

\def \Bu {B_u}

\usepackage[inkscapeformat=png]{svg}
\makeatletter
\newcommand\fs@spaceruled{\def\@fs@cfont{\bfseries}\let\@fs@capt\floatc@ruled
  \def\@fs@pre{\vspace{3mm}\hrule height.8pt depth0pt \kern2pt}%
  \def\@fs@post{\kern2pt\hrule\relax\vspace{-4mm}}%
  \def\@fs@mid{\kern2pt\hrule\kern2pt}%
  \let\@fs@iftopcapt\iftrue}
\makeatother

\title{\LARGE \bf Robust Adaptive MPC Using Uncertainty Compensation*}

\author{Ran Tao$^{1}$, Pan Zhao$^{2}$, Ilya Kolmanovsky$^{3}$, and Naira Hovakimyan$^{1}$
\thanks{*This work is financially supported by Air Force Office of Scientific Research (AFOSR)  Grant \# FA9550-21-1-0411, National Aeronautics and Space Administration (NASA) University Leadership Initiative (ULI) Grant \# 80NSSC17M0051, National Science Foundation's Civil, Mechanical, and Manufacturing Innovation (CMMI)  and Information and Intelligent Systems (IIS) Grants \# 2135925 and \# 2331878, respectively.}
\thanks{$^{1}$R. Tao and N. Hovakimyan are with the Department of 
Mechanical Science and Engineering, University of
Illinois at Urbana-Champaign, Urbana, IL 61801, USA. {\tt\small \{rant3, nhovakim\}@illinois.edu} }%
\thanks{$^{2}$P. Zhao is with the Department of Aerospace Engineering and Mechanics, University of Alabama, Tuscaloosa, AL 35487, USA. {\tt\small pan.zhao@ua.edu} }%
\thanks{$^{3}$I. Kolmanovsky is with the Department of Aerospace Engineering, University of Michigan, Ann Arbor, MI 48109, USA. {\tt\small ilya@umich.edu} }%
\vspace{-5mm}}

\begin{document}

\maketitle
\thispagestyle{empty}
\pagestyle{empty}

\begin{abstract}
This paper presents an uncertainty compensation-based robust adaptive model predictive control (MPC) framework for linear systems with both matched and unmatched nonlinear uncertainties subject to both state and input constraints. In particular, the proposed control framework leverages an $\lone$ adaptive controller (\loneAC) to compensate for the matched uncertainties and to provide guaranteed uniform bounds on the error between the states and control inputs of the actual system and those of a nominal i.e., uncertainty-free, system. The performance bounds provided by the \loneAC~are then used to tighten the state and control constraints of the actual system, and a model predictive controller is designed for the nominal system with the tightened constraints. The proposed control framework, which we denote as uncertainty compensation-based MPC (UC-MPC), guarantees constraint satisfaction and achieves improved performance compared with existing methods. 
Simulation results on a flight control example demonstrate the benefits of the proposed framework.
\end{abstract}



\section{Introduction}\label{sec:introduction}
Real-world systems often require strict adherence to state and/or input constraints stemming from actuator limitations, safety considerations, efficiency requirements, and the need for collision and obstacle avoidance. To deal with systems under constraints, Model Predictive Control (MPC) has emerged as a popular control methodology. MPC optimizes control inputs over a finite time horizon while directly incorporating constraints into the optimization procedure \cite{camacho2013mpc-book,rawlings2020mpc-book}. In addition to the constraints, there are often model uncertainties such as unknown parameters, unmodeled dynamics, and external disturbances, which have to be considered in the control design.
MPC has inherent robustness properties to small uncertainties due to the receding-horizon implementation \cite{yu2014inherent}. However, larger uncertainties need to be explicitly treated in the design. Consequently, researchers have explored {\bf robust or tube MPC} strategies to address uncertainties \cite{kerrigan2001robust-mpc,langson2004robust-mpc-tube,rakovic2005robust-mpc,mayne2006robust-mpc,mayne2011tube-mpc-nonlinear,kohler2020computationally-rmpc,lopez2019dynamic-tube-mpc}, and most of these works consider bounded disturbances. A summary of several robust MPC approaches can be found in \cite{rakovic2018handbook}. However, in practical applications, controllers based on robust MPC can perform conservatively and {\bf adaptive MPC} approaches have also been explored for systems with parametric uncertainties. These approaches typically involve the online identification of a set that is guaranteed to contain the unknown parameters and leverage such a set for robust constraint satisfaction \cite{adetola2011robust,sasfi2023robust,lorenzen2017adaptive,zhang2020adaptive}. However, such approaches require a prior known parametric structure for the uncertainties and, moreover, need the parameters to be time-invariant.

Recently, {\bf uncertainty compensation based constrained control} has also been investigated, which typically involves a control design that estimates and cancels the model uncertainties to force the actual system to behave close to the nominal (uncertainty-free) system. 
For instance, the authors of \cite{zhao2023integrated} integrated an $\lone$ adaptive controller (\loneAC)\cite{hovakimyan2010L1-book} into a reference governor to control systems with time and sate-dependent uncertainties, where the \loneAC~actively compensates for the uncertainties and provides uniform bounds on the error between the states and control inputs of the actual system and a nominal system. Such bounds are then used to tighten the original state and input constraints for robust constraint satisfaction in reference governor design. In \cite{pereida2021robust}, the authors integrated an $\lone$ adaptive controller with robust MPC to deal with systems subject to unknown parameters and disturbances. In particular, \cite{pereida2021robust} considers the state deviation between the actual system and the nominal system from \loneAC~as a bounded model uncertainty and uses robust MPC to handle it. However, the model uncertainty considered in robust MPC is different from the state deviation bound provided by \loneAC~and \cite{pereida2021robust} does not account for the tightening of input constraint resulting from the additional adaptive control input introduced by \loneAC. Additionally, both \cite{zhao2023integrated} and \cite{pereida2021robust} only consider matched uncertainties, i.e., uncertainties injected into the system through the same channels as control inputs. 


\textbf{Contribution:} This paper presents a robust adaptive MPC framework based on uncertainty compensation, which we term as UC-MPC, for linear systems with nonlinear and time-varying uncertainties that can contain both matched and unmatched components. Our framework leverages an \loneAC~to estimate and compensate for the matched uncertainties, and to guarantee uniform bounds on the error between states and inputs of the actual system and those of a nominal (i.e., uncertainty-free) closed-loop system. Such bounds are then used to tighten the original constraints. The tightened constraints are leveraged in the design of MPC for the nominal system, which guarantees robust constraint satisfaction of the actual system in the presence of uncertainties. To validate the effectiveness of our framework, we conduct simulations using a flight control example. 

The proposed UC-MPC has the following features: 
\begin{itemize}
    \item In addition to the enforcement of the constraints, UC-MPC improves the tracking performance compared with existing robust or tube MPC solutions \cite{kerrigan2001robust-mpc,langson2004robust-mpc-tube,rakovic2005robust-mpc,mayne2006robust-mpc,mayne2011tube-mpc-nonlinear,kohler2020computationally-rmpc,lopez2019dynamic-tube-mpc}, due to the active uncertainty compensation. 
    \item UC-MPC handles a broad class of uncertainties that can be time-varying and state-dependent without a parametric structure, while existing adaptive MPC approaches\cite{adetola2011robust,sasfi2023robust,lorenzen2017adaptive,zhang2020adaptive} typically handle parametric uncertainties only.
    \item UC-MPC can also handle unmatched disturbances while existing uncertainty compensation-based constrained control only considers matched uncertainties \cite{pereida2021robust,zhao2023integrated}.
\end{itemize}


{\it Notations}: In this paper, we use $\mathbb{R}$, $\mathbb{R}_+$ and $\mbZ_+$ to denote the set of real, non-negative real, and non-negative integer numbers, respectively. $\mathbb{R}^n$ and  $\mathbb{R}^{m\times n}$ represent the $n$-dimensional real vector space and the set of real $m$ by $n$ matrices, respectively. $\mbZ_i$ and $\mbZ_1^n$ denote the integer sets $\{i, i+1, \cdots\}$ and $\{1, 2,\cdots,n\}$, respectively.
$I_n$ denotes a size $n$ identity matrix, and $0$ is a zero matrix of a compatible dimension.
$\norm{\cdot}$ and $\norm{\cdot}_\infty$ represent the $2$-norm and $\infty$-norm of a vector or a matrix, respectively. 
 The $\mathcal{L}_\infty$- and truncated $\mathcal{L}_\infty$-norm of a function $x:\mathbb{R}_+ \rightarrow\mathbb{R}^n$ are defined as $\norm{x}_{\mathcal{L}_\infty}\triangleq \sup_{t\geq 0}\infnorm{x(t)}$ and $\linfnormtruc{x}{T}\triangleq \sup_{0\leq t\leq T}\infnorm{x(t)}$, respectively. The Laplace transform of a function $x(t)$ is denoted by $x(s)\triangleq\mathfrak{L}[x(t)]$.
 Given a vector $x$, $x_i$ denotes the $i$th element of $x$. For positive scalar $\rho$, $\Omega(\rho)\trieq \{z\in \mbR^n: \infnorm{z}\leq \rho \}$ represents a high dimensional ball set of radius $\rho$ which centers at the origin with a compatible dimension $n$. For a high-dimensional set $\mcX$, the interior of $\mcX$ is denoted by $\textup{int}(\mcX)$ and the projection of $\mcX$ onto the $i$th coordinate is represented by $\mcX_i$. For given sets $\mcX,\mcY\subset \mbR^n$, $\mcX \oplus \mcY \trieq \{ x+y: x\in \mcX, y\in \mcY \}$ is the Minkowski set sum and $\mcX\ominus \mcY \trieq \{z: z+y\in \mcX, \forall y\in \mcY \}$ is the Pontryagin set difference. 
\section{Problem Statement}
Consider a linear system with uncertainties represented by
\begin{equation}\label{eq:dynamics-uncertain-original}
\left\{ \begin{aligned}
  \dot x(t) &= Ax(t) + B(u(t) +  f(t,x(t)))+B_u w(t), \hfill \\
  y(t) &= Cx(t), \ x(0) = x_0,\\ 
\end{aligned}\right.    
\end{equation}
where $x(t)\in\mbR^n$, $u(t)\in \mbR^m$ and $y(t)\in\mbR^m$ represent the state, input, and output vectors, respectively, $x_0\in\mbR^n$ is the initial state, and matrices $A$, $B$, $B_u$, and $C$ are known with compatible dimensions. $B$ has full column rank, and $B_u$ is a matrix such that $\textup{rank}[B\  B_u]= n$  and $B_u^T B =0 $. Moreover, $f(t,x(t))\in \mbR^m$ represents the matched uncertainty dependent on both time and states, and $w(t)\in \mbR$ denotes the unmatched uncertainty. 
\begin{assumption}\label{assump:lipschitz-bnd-fi}
Given a compact set $\mcZ$, there exist positive constants $L_{f_j,\mcZ}$, $l_{f_j,\mcZ}$, and $b_{f_j,\mcZ}$ ($j\in\Zm$) such that for any $x,z \in \mcZ$ and $t,\tau\geq 0$, the following inequalities hold for all $j\in\Zm$:
\begin{subequations}\label{eq:lipschitz-cond-and-bnd-fi}
\begin{align}
\abs{f_j(t,x) - f_j(\tau,z)}  & \!\le\! L_{f_j,\mcZ}\infnorm{x - z} \!+\! l_{f_j,\mcZ}\abs{t-\tau} , \label{eq:lipschitz-cond-fi}\\
  \abs {f_j(t,x)}  &\le b_{f_j,\mcZ}, \label{eq:bnd-fi} 
\end{align}
\end{subequations}
where $f_j(t,x)$ denotes the $i$th element of $f(t,x)$. In addition, there exist a positive constant $b_{w}$ such that for any $t\geq 0$, we have:
\begin{equation}\label{eq:bnd-wi}
      \infnorm {w(t)}  \le b_{w}.
\end{equation}
\end{assumption}  
\begin{remark}
    The function $f(t,x)$ represents the matched uncertainty that can be directly canceled using control inputs. On the other hand, $w(t)$ denotes the unmatched uncertainty that cannot be directly canceled. Since the proposed solution needs to be robust against the unmatched uncertainty, we restrict $w(t)$ to be bounded and state-independent to simplify the derivation. It is possible to slightly extend the results to allow $w(t)$ to be state-dependent. 
\end{remark}
Based on \cref{assump:lipschitz-bnd-fi}, it follows that for any $x,z \in \mcZ$ and $t,\tau\geq 0$, we have
\begin{subequations}\label{eq:lipschitz-cond-bnd-f}
\begin{align}
\infnorm{f(t,x) - f(\tau,z)}  & \!\le\!{L_{f,\mcZ}}\infnorm{x - z} \!+\! l_{f,\mcZ}\abs{t-\tau}, \label{eq:lipschitz-cond-f}\\
  \infnorm {f(t,x)}  &\le {b_{f,\mcZ}},  \label{eq:bnd-f} \\
  \infnorm {w(t)}  &\le b_{w}, \label{eq:bnd-w} 
\end{align}
\end{subequations}
where 
\begin{align}
\label{eq:Lf-lf-bf-defn}
        L_{f,\mcZ} &= \max_{j\in \Zm} L_{f_j,\mcZ},\; l_{f,\mcZ} = \max_{j\in \Zm} l_{f_j,\mcZ},\nonumber \\ 
        b_{f,\mcZ} &= \max_{j\in \Zm} b_{f_j,\mcZ}.
\end{align}

\begin{remark}
We specifically make assumptions on $f_j(t,x)$ instead of on $f(t,x)$ as in \cref{eq:Lf-lf-bf-defn} in order to derive an  {\it individual} bound on each state and on each adaptive input (see \cref{sec:l1acbounds} for details).
\end{remark}

Given the system \cref{eq:dynamics-uncertain-original}, the goal of this paper is to design an MPC scheme with active uncertainty compensation to achieve desired performance while satisfying the following state and control constraints:
\begin{equation}\label{eq:constraints}
    \begin{gathered}
  x(t) \in \mathcal{X},\quad u(t) \in \mathcal{U}, \quad \forall t\geq 0,\hfill
\end{gathered} 
\end{equation}
where $\mcX\subset \mbR^n$ and $\mcU \subset\mbR^m$ are pre-specified convex and compact sets including the origin. 

\section{Overview of the UC-MPC Framework}\label{sec:overview}
\begin{figure}[!b]
     \centering
     \includegraphics[width = \columnwidth]{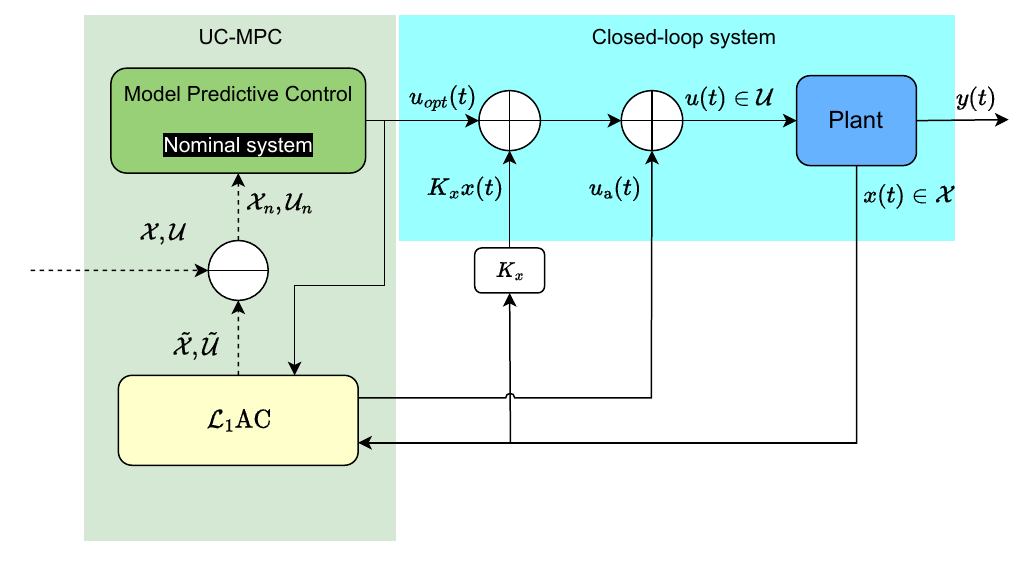}
     \caption{Diagram of the proposed UC-MPC framework}
     \label{fig:framework}
     \vspace{-3mm}
 \end{figure}
The schematic diagram of the proposed UC-MPC is shown in \cref{fig:framework}, which includes a feedback control law $K_x x(t)$, an \loneAC~to generate $u_a(t)$, and an MPC design for the nominal system with tightened constraints $\mcX_\nt$ and $\mcU_\nt$ to generate $u_\text{opt}(t)$.
In the presence of the uncertainties $f(t,x)$ and $w(t)$, we leverage an \loneAC~to generate $u_a(t)$ to compensate for the matched uncertainty and to force the actual system to behave close to a nominal (i.e., uncertainty-free) system \cite{hovakimyan2010L1-book}. 
The feedback control law $K_xx(t)$ ensures that $A+BK_x$ is Hurwitz. 
Thus, the overall control law for system \cref{eq:dynamics-uncertain-original} can be represented as:
\begin{equation}\label{eq:total-control-law}
    u(t)=K_xx(t)+u_\text{opt}(t)+u_a(t).
\end{equation}
With \cref{eq:total-control-law}, we can reformulate the system \cref{eq:dynamics-uncertain-original} as
\begin{equation}\label{eq:dynamics-uncertain}
\left\{
\begin{aligned}
 \!\dot x(t) \!&=\! {A_m}x(t)\! + \!Bu_\text{opt}(t) \!+\! B(u_\at(t) \!+\! f(t,\!x(t)\!)\!)\!+\!B_uw(t) \\ 
 \! y(t) & = Cx(t), \ x(0) =x_0, 
\end{aligned}\right. 
\end{equation}
where ${A_m} \trieq A + B{K_x}$ is a Hurwitz matrix. In \cref{sec:l1acbounds}, we will show that the existence of \loneAC~provides uniform bounds on the errors between the states and control inputs of the real plant \cref{eq:dynamics-uncertain} and those of the nominal system represented by 
\begin{equation}\label{eq:nominal system}
\left\{
\begin{aligned}
  \dot x_\nt(t) &= {A_m}\xn(t) + Bu_\text{opt}(t), \ \xn(0) =x_0, \hfill \\
     u_\nt(t)& =K_x \xn(t) + u_\text{opt}(t),
\end{aligned}\right.
\end{equation}
where $x_\nt$ and $u_\nt$ are the vectors of nominal states and inputs. Using the uniform bounds from \loneAC, we design an MPC for the nominal system with tightened constraints. In particular, from \loneAC, we achieve 
\begin{equation}\label{eq:x-xn-in-X-u-un-in-U}
    \begin{aligned}
    x(t)-\xn(t) \in \tilde \mcX,  
  \quad  u(t)- \un(t) \in  \tilde \mcU, \quad \forall t\geq 0,
    \end{aligned}
\end{equation}
where $u(t)$ is defined in \cref{eq:total-control-law} and $\un(t)$ is given in \cref{eq:nominal system},
and $\tilde \mcX$ and $\tilde \mcU$ are some pre-computed hyperrectangular sets calculated from the range of $f(t,x)$ and $w(t)$ and the design of \loneAC.
Given the state and control input constraints of the real plant $\mathcal{X}$ and $\mathcal{U}$ in \cref{eq:constraints}, the tightened constraints for the nominal system \cref{eq:nominal system} are represented by 
\begin{equation}\label{eq:Xn-Un-defn}
    \mcX_\nt \trieq \mcX \ominus \tilde \mcX, \quad \mcU _\nt \trieq \mcU \ominus \tilde \mcU.
\end{equation}


\section{\loneAC~With Uniform Performance Bounds}\label{sec:l1acbounds}
We now present the uniform performance bounds provided by an \loneAC~on the errors between the states and control inputs of the real plant \cref{eq:dynamics-uncertain} and those of the nominal system \cref{eq:nominal system}. A preliminary result is presented in \cite{zhao2023integrated}, in which the authors considered only matched uncertainties. In this paper, we extend the result to systems with both matched uncertainty $f(t,x(t))$ and unmatched uncertainty $w(t)$. 

An \loneAC~usually includes three elements: a state predictor, an estimation law, and a low-pass filter. 
The inclusion of a low-pass filter $\mcC(s)$ (with DC gain $\mcC(0)=I_m$) decouples the estimation loop from the control loop, which enables fast adaptation without sacrificing the robustness \cite{hovakimyan2010L1-book}. The {\bf low-pass filter} $\mcC(s)$ can be designed as a first-order transfer function matrix
\begin{equation}\label{eq:filter-defn}
\mcC(s) = \textup{diag}(\mcC_1(s), \dots, \mcC_m(s)), \ \mcC_j(s) \trieq \frac{k_f^j}{(s+ k_f^j)},
\end{equation}
where  $k^j_f$ ($j\in\Zm$) represents the bandwidth for the $j$th input channel. 
In order to ensure stability, given the compact set $\mcX_0$, the filter $\mcC(s)$ defined in \cref{eq:filter-defn} needs to ensure that there exists a positive constant $\rho_r$ and a (small) positive constant $\gamma_1$ such that
\begin{subequations}\label{eq:l1-stability-condition-w-extra-Lf}
\begin{align}
\lonenorm{\mcG_{xm} (s )} b_{f,\mcX_r}  & <  \rho_r   -  \lonenorm{\mcH_{xm}(s)} \linfnorm{u_\text{opt}} \nonumber \\&- \lonenorm{\mcH_{xu}(s)}b_w - \rho_\textup{in},\label{eq:l1-stability-condition} \\ 
\lonenorm{\mcG_{xm}(s)}L_{f,\mcX_a}&<1,\label{eq:l1-stability-condition-Lf}
\end{align}
\end{subequations}
where 
\begin{align} 
 \rho  & \trieq \rho_r + \gamma_1, \label{eq:rho-defn} \\
 \mcX_r & \trieq \Omega(\rho_r), \   \mcX_a  \trieq  \Omega(\rho). \label{eq:Xr-Xa-defn}\\
 \mcH_{xm}(s) & \trieq\! (sI_n \!-\! A_m)^{-1}\! B,\label{eq:Hxm_defn}\\
 \mcH_{xu}(s) &\trieq\! (sI_n \!-\! A_m)^{-1} \!\Bu, \label{eq:Hxu-defn}\\
 \mcG_{xm}(s) & \trieq\! \mcH_{xm}(s)(I_m-\mcC(s)),
   \label{eq:Gm-defn} \\
 \rho_\textup{in} &\trieq \lonenorm{s(sI_n-A_m)^{-1}}\max_{x_0\in \mcX_0}\infnorm{x_0}. \label{eq:rho_in}
\end{align}
Also, we define $\xin(t)$ as the state of the system 
    $\dot{x}_\textup{in}(t) = A_m \xin(t), \  \xin(0) = x_0,$
and we have $\xin(s) \trieq (sI_n-A_m)^{-1} x_0$. 

For the real system \cref{eq:dynamics-uncertain}, the {\bf state predictor} is defined as 
\begin{multline}\label{eq:state-predictor}
\dot {\hat x} (t)\!=\! A_m x(t) \! +\! B u_\text{opt}(t) \!+\! B u_\at (t)\! +\! \hsigma(t) \!+\!  A_e \tilx(t),
\end{multline}
where $\hat x(0) = x_0$, $\tilx(t) = \hat x(t) - x(t)$ is the prediction error, $A_e$ is a custom Hurwitz matrix, and $\hsigma (t)$ is the estimate of the lumped uncertainty, $Bf(t,x)+B_uw(t)$. The estimate $\hsigma(t)$ is updated according to the following piecewise-constant {\bf estimation law} (similar to that in \cite[Section~3.3]{hovakimyan2010L1-book}):
\begin{equation}\label{eq:adaptive_law}\left\{
\begin{aligned}
&  \hsigma(t) &  &  \hspace{-4mm}=  
\hsigma(iT)
    , \quad t\in [iT, (i+1)T), \\
& \hsigma(iT) &
& \hspace{-4mm}= - \Phi^{-1}(T)e^{A_eT}\tilde{x}(iT),
\end{aligned}\right.
\end{equation} where $T$ is the estimation sampling time, and $\Phi(T)\trieq A_e^{-1}\left(e^{A_eT}\!-I_n\right)$. The control law of the adaptive controller is given by 
\begin{equation}\label{eq:l1-control-law}
   u_\at(s) = -\mcC(s)\laplace{B^\dagger \hsigma(t)},
\end{equation}
where $B^\dagger=(B^TB)^{-1}B^T$ is the pseudo-inverse of $B$. The control law \cref{eq:l1-control-law} tries to cancel the estimate of the matched uncertainty $f(t,x(t))$ within the bandwidth of the filter $\mcC(s)$. 

Now we define some constants below for future reference:
\begin{subequations}\label{eq:alpha_012-gamma0-T-defn}
\begin{align}
    \bar \alpha_0(T)  \trieq& \max_{t\in [0,T]} \int_0^t\infnorm{e^{A_e(t-\tau)}B}d\tau,\label{eq:alpha_0-defn}\\
    \bar \alpha_1(T)  \trieq& \max_{t\in [0,T]}\int_0^t\infnorm{e^{A_e(t-\tau)}B_u}d\tau,\label{eq:alpha_1-defn} \\
    \bar \alpha_2(T) \trieq& \max_{t\in [0,T]} \infnorm{e^{A_e t}},
   \label{eq:alpha_2-defn} \\
      \bar \alpha_3(T) \trieq& \max_{t\in [0,T]}\int_0^t \infnorm{e^{A_e (t-\tau)}\Phi^{-1}(T)e^{A_eT}}d\tau,\label{eq:alpha_3-defn}
         \\
        \gamma_0(T) \trieq& (b_{f,\mcX_a} \bar\alpha_0(T)\!+\!\bar \alpha_1(T)b_w)\!(\bar \alpha_2(T) \!+\! \bar \alpha_3(T)\!+\!1\!)\!\label{eq:gamma_0-T-defn} 
        \\ \hspace{-4mm}    \rho_{ur} \trieq&  \lonenorm{\mcC(s)} b_{f,\mcX_r}, \label{eq:rho_ur-defn} \\
\hspace{-2mm}  \gamma_2  \trieq &  \lonenorm{\mcC(s)}\! L_{f,\mcX_a} \gamma_1 \!+\! \lonenorm{\mcC(s)B^\dagger(sI_n\!-\!A_e)}\!\gamma_0(T), \label{eq:gamma2-defn} \\
 \hspace{-4mm}  \rho_{u_\at} \trieq&  \rho_{ur} +\gamma_2, \label{eq:rho-u-defn}
\end{align}
\end{subequations}
where $\gamma_1$ is introduced in \cref{eq:rho-defn}. Based on the Taylor series expansion of $e^{A_eT}$, we have $\lim_{T\rightarrow 0} \int_0^T \infnorm{\Phi^{-1}(T)}d\tau$ is bounded, which further implies that $ \lim_{T\rightarrow 0} \bar \alpha_3(T)$ is bounded. Since $\lim_{T\rightarrow 0}\bar \alpha_0(T) =0 $, $\lim_{T\rightarrow 0}\bar \alpha_1(T) =0$, $\lim_{T\rightarrow 0} \bar \alpha_2(T)=1$, $b_{f,\mcX_a}$ is bounded for a compact set $\mcX_a$, $b_w$ is bounded, and $\lim_{T\rightarrow 0} \bar \alpha_3(T)$ is bounded, we have
\begin{equation}\label{eq:gammaT-to-0}
    \lim_{T\rightarrow 0} \gamma_0(T) = 0.
\end{equation}
Considering \cref{eq:gammaT-to-0,eq:l1-stability-condition-Lf}, it is always feasible to find a small enough $T>0$ such that
\begin{equation}\label{eq:T-constraint}
\frac{\lonenorm{\mcH_{xm}(s)\mcC(s)B^\dagger(sI_n-A_e)}}{1-\lonenorm{\mcG_{xm}(s)}L_{f,\mcX_a}}\gamma_0(T) < \gamma_1,
\end{equation}
 where $\mcX_a$ is defined in \cref{eq:Xr-Xa-defn} and $B^\dagger$ is the pseudo-inverse of $B$. 

Following the convention of an \loneAC \cite{hovakimyan2010L1-book}, we introduce the following reference system for performance analysis: 
\begin{align}
  \dot{x}_\rt(t) &\! =\! {A_m}x_\textup{r}(t) \!\!+\!\! {B}u_\text{opt}(t) \!\!+\! \!B({u_{\textup{r}}}(t)\! \!+\!\! f(t,\!x_\textup{r}(t)))\!\!+\!\!B_uw(t), \hfill \nonumber   \\
  {u_{\textup{r}}}(s) & \! =\! -\mcC(s)\laplace{f(t,x_\textup{r}(t))}, \quad\xr(0)\! =\!x_0, \label{eq:ref-system} 
\end{align}
The control law in \eqref{eq:ref-system} depends on the true
uncertainties and is thus {\it not implementable}. The reference system is introduced to connect the adaptive closed-loop system with the nominal system and help characterize the performance of the adaptive closed-loop system. \vspace{-1mm}

\begin{lemma}\label{them:x-xref-bnd}
{Given the uncertain system \cref{eq:dynamics-uncertain} subject to  \cref{assump:lipschitz-bnd-fi} and  the reference system \cref{eq:ref-system} subject to the conditions \cref{eq:l1-stability-condition-Lf,eq:l1-stability-condition}  with a constant $\gamma_1>0$, with the \loneAC~defined via \cref{eq:state-predictor,eq:adaptive_law,eq:l1-control-law} subject to the sample time constraint  \cref{eq:T-constraint}, we have }
\begin{subequations}
\begin{align}
    \linfnorm{x} & \leq \rho, \label{eq:x-bnd}\\
    \linfnorm{u_\at} & \leq \rho_{u_\at}, \label{eq:ua-bnd}\\
    \linfnorm{x_\textup{r}-x} &\leq \gamma_1, \label{eq:xref-x-bnd}\\
    \linfnorm{u_{\textup{r}}-u_\at} &\leq \gamma_2, \label{eq:uref-u-bnd}
\end{align}
\end{subequations}
where $\rho$, $\rho_{u_\at}$, and $\gamma_2$ are defined in \cref{eq:rho-defn}, \cref{eq:rho-u-defn}, \cref{eq:gamma2-defn}, respectively. 
\end{lemma}

\begin{lemma} \label{lem:ref-id-bnd}
Given the reference system \cref{eq:ref-system} and the nominal system \cref{eq:nominal system},  subject to \cref{assump:lipschitz-bnd-fi}, and the condition  \cref{eq:l1-stability-condition}, we have
\begin{align}
{\left\| {{x_\textup{r}} - \xn} \right\|_{{\mathcal L}{_\infty }}} & \leq  \lonenorm{\mcG_{xm}}b_{f,\mcX_r} +\lonenorm{\mcH_{xu}}b_w \label{eq:xref-xid-bnd} .
\end{align}
\end{lemma}
\cref{them:x-xref-bnd} and \cref{lem:ref-id-bnd} are relatively straightforward extensions of \cite[Theorem~1 and Lemma~5]{zhao2023integrated} with the consideration of the unmatched uncertainty $w(t)$. The proofs can be obtained by extending the proofs of \cite[Theorem~1 and Lemma~5]{zhao2023integrated} and are included in the appendix.

The previous results provide uniform error bounds as represented by the vector-$\infty$ norm, which always leads to the {\it same bound} for all the states, $x_i-x_{\nt,i}(t)$ ($i\in\mbZ_1^n$), or all the adaptive inputs, $u_{\at,j}$ ($j\in\mbZ_1^m$). The use of vector-$\infty$ norms may lead to conservative bounds given some states or adaptive inputs, making it impossible to satisfy the constraints \cref{eq:constraints} or leading to significantly tightened constraints for the MPC design. Thus, an individual bound for each  $x_i(t)-x_{\textup r,i}(t)$ is preferred.
To derive the individual bounds for each $i\in\Zn$, we follow \cite{zhao2023integrated} to introduce the following coordinate transformations for the reference system \cref{eq:ref-system} and the nominal system \cref{eq:nominal system}:
\begin{equation}\label{eq:coordinate-trans}
\left \{
    \begin{aligned}
        \check x_\rt & = \Tau_x^ix_\rt,\quad \check x_{\nt} = \Tau_x^i \xn, \\
         \check A_m^i &= \Tau_x^iA_m (\Tau_x^i)^{-1},\\
         \check B^i  & = \Tau_x^iB,   \quad \check B^i _u = \Tau_x^iB_u,
    \end{aligned}
    \right.
\end{equation}
where $\Tau_x^i\!>\!0$ is a diagonal matrix that satisfies 
\begin{align}
    \Tau_x^i[i]&=1, \ 0<\Tau_x^i[k]\leq 1, \ \forall k\neq i, \label{eq:Tx-i-cts}
\end{align}
and $\Tau_x^i[k]$ is the $k$th diagonal element. With the transformation \cref{eq:coordinate-trans}, the reference system \cref{eq:ref-system} is transformed into 
\begin{align}
  \dot {\check x}_\rt(t) & \!=\! \check A_m^i\check x_\rt(t)\!\! +\!\!  \check  B^iu_\text{opt}(t) \!\!+\!\!  \check B^i (u_{\rt}(t) \!\!+\!\! \check f(t, \check x_\rt(t)\!)\!)\!+\!\check B_u^iw(t),  \nonumber\\ 
    {u_{\textup{r}}}(s) & \! =\! -\mcC(s)\laplace{\check f(t,\check x_\textup{r}(t))},  \ \check x(0) \!=\! \Tau_x^ix_0,\label{eq:ref-system-transformed}
\end{align}
where 
\begin{equation}\label{eq:f-checkf-relation}
    \check f(t,\check x_\rt(t)) =  f(t, x_\rt(t))) =  f(t, (\Tau_x^i)^{-1}\check x_\rt(t))).
\end{equation}Given a set $\mcZ$, define 
\begin{equation}\label{eq:check-Z-defn}
    \check \mcZ\trieq \{\check z\in \mbR^n: (\Tau_x^i)^{-1}\check z \in \mcZ\}.
\end{equation}
For the transformed reference system \cref{eq:ref-system-transformed}, we have
\begin{subequations}\label{eq:Hxm-Hxv-Gxm-check-defn}
\begin{align}
  \mcH_{\check xm}^i(s) &\trieq (sI_n \!-\! \check A_m^i)^{-1}\! \check B^i  = \Tau_x^i \mcH_{xm}(s), \label{eq:Hxm-check-defn} \\
  \mcH_{\check xu}^i(s) & \trieq (sI_n \!-\! \check A_m^i)^{-1} \! \check B^i _u = \Tau_x^i \mcH_{xu}(s), \label{eq:Hxv-check-defn} \\
  \mcG_{\check xm}^i(s) & \trieq \mcH_{\check xm}^i(s)(I_m-\mcC(s)) =  \Tau_x^i \mcG_{xm}(s), \label{eq:Gxm-check-defn}
\end{align}
\end{subequations}
where $\mcH_{xm},~\mcH_{xu},~\mcG_{xm}$ are defined in \cref{eq:Hxm_defn,eq:Hxu-defn,eq:Gm-defn}. 
By applying the transformation \cref{eq:coordinate-trans} to the nominal system \cref{eq:nominal system}, we obtain 
\begin{equation}\label{eq:nominal-cl-system-transformed}
\hspace{-2mm}
\left\{
\begin{aligned}
  \dot {\check x}_{\textup n}(t) & = \check A_m^i\xcheckn(t) + \check B^iu_\text{opt}(t),\ \xcheckn(0) \!=\! \Tau_x^ix_0,  \\ 
  \check y_\text{n}(t) & = \check C \xcheckn(t).
\end{aligned}\right. 
\end{equation}
Letting $\xcheckin(t)$ be the state of the system 
    $\dot{\check x}_\textup{in}(t) = \check A_m^i \xcheckin(t)$ with $\xcheckin(0) = \check x_\nt(0) = \Tau_x^i x_0$, 
we have $\xcheckin(s) \trieq (sI_n-\check A_m^i)^{-1} \xcheckin(0) = \Tau_x^i(sI_n- A_m)^{-1} x_0$. Define
\begin{equation}\label{eq:check-rhoin-defn}
   \check \rho_\textup{in}^i \trieq \lonenorm{s\Tau_x^i(sI_n- A_m)^{-1}}\max_{x_0\in \mcX_0}\infnorm{x_0}.
\end{equation}
Similar to \cref{eq:l1-stability-condition}, for the transformed reference system \cref{eq:ref-system-transformed}, given any positive constant $\gamma_1$, the lowpass filter design now needs to satisfy:
\begin{align}
\lonenorm{\mcG_{\check xm}^i(s)}b_{\check f,\check \mcX_r}  &< \check \rho_r^i - \lonenorm{\mcH_{\check x v}^i(s)}\linfnorm{u_\text{opt}} \nonumber\\&- \lonenorm{\mcH_{\check x u}^i(s)}b_w - \check \rho_\textup{in}^i ,\label{eq:l1-stability-condition-transformed} %
\end{align}
where 
$\mcX_r$ is defined in \cref{eq:Xr-Xa-defn} and $\check \mcX_r$ is defined according to \cref{eq:check-Z-defn}, and $\check \rho_r^i$ is a positive constant to be determined. 


\begin{lemma}\label{lem:refine_bnd_xi_w_Txi}
Consider the reference system \cref{eq:ref-system} 
subject to \cref{assump:lipschitz-bnd-fi}, the nominal system \cref{eq:nominal system}, the transformed reference system \cref{eq:ref-system-transformed} and transformed nominal system \cref{eq:nominal-cl-system-transformed} obtained by applying \cref{eq:coordinate-trans} with any $\Tau_x^i$ satisfying \cref{eq:Tx-i-cts}. 
Suppose that \cref{eq:l1-stability-condition} holds with some constants $\rho_r$ and $\linfnorm{u_\text{opt}}$.
Then, 
there exists a constant $\check \rho_r^i\leq \rho_r$ such that \cref{eq:l1-stability-condition-transformed}
holds with the same $\linfnorm{u_\text{opt}}$. Furthermore, $\forall t\geq 0$, 
\begin{align}
\abs{x_{\rt,i}(t)} & \leq  \check \rho_r^i, \label{eq:xr-i-bnd-from-trans} \\
    \abs{x_\textup{r,i}(t)-x_{\nt,i}(t)} & \leq \lonenorm{\mcG_{ \check xm}(s)}b_{f,\mcX_r} \!+\! \lonenorm{\mcH_{ \check xu}} b_w, \label{eq:xri-xni-bnd-from-trans}
\end{align}
where we re-define \begin{equation}\label{eq:Xr-defn}
   \mcX_r\trieq \left\{z\in\mbR^n: \abs{z_i}\leq \check \rho_r^i, i\in\Zn\right\}. 
\end{equation}
\end{lemma}

\begin{lemma}\label{them:xi-uai-bnd}
Consider the uncertain system \cref{eq:dynamics-uncertain} subject to \cref{assump:lipschitz-bnd-fi}, the nominal system \cref{eq:nominal system}, and the \loneAC~defined via \cref{eq:state-predictor,eq:adaptive_law,eq:l1-control-law} subject to the conditions \cref{eq:l1-stability-condition-Lf,eq:l1-stability-condition} with constants $\rho_r$ and $\gamma_1>0$ and the sample time constraint \cref{eq:T-constraint}. Suppose that for each $i\in \Zn$, 
\cref{eq:l1-stability-condition-transformed} holds with a constant $\check \rho_r^i$ for the transformed reference system \cref{eq:ref-system-transformed} obtained by applying \cref{eq:coordinate-trans}. Then, $\forall t\ge0$, we have 
\begin{subequations}\label{eq:xui-xuni-bnd-from-trans-w-tilX-tilU-defn}
\begin{align}
\hspace{-2mm}     {x(t)-x_{\nt}(t)} \!\in\! \tilde \mcX\! &\trieq\! \left\{z\!\in\!\mbR^n\!: \!\abs{z_i}\!\leq\! \tilde \rho^i, \  i\in \Zn \right\}\!,\ \label{eq:xi-xni-bnd-from-trans-w-tilX-defn}\\
 \hspace{-2mm} {u_{\at}(t)} \!\in\! \mcU_\at  \!&\trieq\! \left\{z\!\in\!\mbR^m\!: \!\abs{z_j}\!\leq\! \rho_{u_\at}^j, \ \! j\!\in\! \Zm \right \}\!, \label{eq:ua-i-bnd-w-Ua-defn} \\ 
 \hspace{-2mm}        u(t) - u_{\nt}(t)\!\in\!  \tilde\mcU   \!&\trieq\! \left \{z \!\in\! \mbR^m\!: \!\abs{z_j}\!\leq\!  \tilde\rho_u^j,
\  j\!\in\! \Zm \right \}\!, \label{eq:uj-unj-bnd-w-tilU-defn} 
\end{align}
\end{subequations}where 
\begin{subequations}\label{eq:rhoi-tilrhoi-y-defn}
\begin{align}
 \hspace{-3mm}   &   \rho^i   \!\trieq\! \check \rho_r^i\!+\!\gamma_1, \tilde \rho^i\!\trieq\! \lonenorm{\mcG_{ \check xm}^i(s)}\!\!b_{f,\mcX_r}\!+\!\lonenorm{\mcH^i_{ \check xu}} \!\!b_w +\!\gamma_1,\!\label{eq:rhoi-tilrhoi-defn} 
\\
\hspace{-3mm} & \rho_{u_\at}^j  \!\trieq\! \lonenorm{\mcC_j(s)} b_{f_j, \mcX_r}\! +\! \gamma_2,\!   \tilde\rho_u^j \!\trieq \!  \rho_{u_\at}^j \!+\!
\sum_{i=1}^n\abs{K_x[j,i]}\tilde \rho^i
\label{eq:tilrho-u-j-defn}
\end{align}
\end{subequations}
with $\mcX_r$ defined in \cref{eq:Xr-defn}, and $C[j,i]$ denoting the $(j,i)$ element of $C$.
\end{lemma}
\cref{lem:refine_bnd_xi_w_Txi} and \cref{them:xi-uai-bnd} are the extensions of \cite[Lemma 6 and Theorem 3]{zhao2023integrated} to account for the unmatched uncertainty $w(t)$. The proofs can be found in the appendix.

\begin{remark}\label{rem:x-xid-discussion-Cs-T-refined}
\cref{them:xi-uai-bnd} provides a method to derive an individual bound on $x_i(t)-x_{\nt,i}(t)$ for each $i\in \Zn$ and on $u_j (t) - u_{\nt,j}(t)$ for each $j\in\Zm$ via coordinate transformations.
Additionally, by decreasing $T$ and increasing the bandwidth of the filter $\mcC(s)$, one can make $\tilde \rho^i$ ($i\in \Zn$) arbitrarily small, i.e., making {\it the states of the adaptive system arbitrarily close to those of the nominal system}, in the absence of the unmatched uncertainty, and make the bounds on $u_{\at,j}(t)$ and $u_j(t)-u_{\nt,j}(t)$ arbitrarily close to the bound on the true matched uncertainty $f_j(t,x)$ for $x\in\mcX_a$, and for each $j\in\Zm$.
\end{remark}

\section{UC-MPC: Robust MPC via Uncertainty Compensation}
According to \cref{them:xi-uai-bnd}, the procedure for designing the \loneAC~to compensate for the uncertainties and the tightened bounds for the nominal system can be summarized in \cref{al1}. 

It is worth mentioning that we additionally constrain $\xr(t)$ and $x(t)$  to stay in $\mcX$ for all $t\geq0$
in step~\ref{line:l1-stability-conditions-in-l1rg} of \cref{al1}. 
Such constraints 
can potentially reduce the uncertainty size that needs to be compensated and significantly reduce the conservatism of the proposed method. 

\begin{algorithm}
\caption{UC-MPC Design}\label{al1}
\begin{algorithmic}[1]
\Require{Uncertain system \cref{eq:dynamics-uncertain-original} subject to \cref{assump:lipschitz-bnd-fi} with constraint sets $\mcX$ and $\mcU$ defined in \cref{eq:constraints}, set for the initial condition $\mcX_0$, $A_e$ for state predictor \cref{eq:state-predictor}, initial low pass filter $\mcC(s)$ and sampling time $T$ to define an \loneAC, $\gamma_1$, tol, and matrix $K_x$}
\State For the nominal system defined in \cref{eq:nominal system}, find the maximum possible range of $u_\text{opt}$ and calculate $\linfnorm{u_\text{opt}}$ when the constraints are given by $x_\nt\in \mcX$ and $u_\nt\in \mcU$

\While{\cref{eq:l1-stability-condition} with $\mcX_r = \Omega(\rho_r) \cap \mcX$ or \cref{eq:l1-stability-condition-Lf} with $\mcX_a = \Omega(\rho_r+\gamma_1) \cap \mcX$
does not hold with any $\rho_r$}\label{line:l1-stability-conditions-in-l1rg}
\State Increase the bandwidth of $\mcC(s)$
\EndWhile \Comment{$\rho_r$, $\mcX_r$ and $b_{f,\mcX_r}$  will be computed.} \label{line:Xr-b_f-under-csts}
\State Set $b_{f,\mcX_r}^{old} = b_{f,\mcX_r}$
\For{$i=1,\dots,n$}\label{line:Ti}
\State Select $\Tau_x^i$ satisfying\!~\cref{eq:Tx-i-cts} and apply the transformation\!~\cref{eq:coordinate-trans}\label{line:Tx_i-selection} 
\State Evaluate \cref{eq:Hxm-Hxv-Gxm-check-defn} and 
compute $\check \rho_\textup{in}^i$ according to \cref{eq:check-rhoin-defn}
\State Compute $\check \rho_{r}^i$ that satisfies \cref{eq:l1-stability-condition-transformed}
\State Set $\rho^i =  \check \rho_r^i+\gamma_1$, $\tilde \rho^i= \lonenorm{\mcG_{ \check xm}^i(s)}\!\!b_{f,\mcX_r}\!+\!\lonenorm{\mcH^i_{ \check xu}} \!\!b_w +\!\gamma_1$
\EndFor

\State Set $\mcX_r\! = \!\left\{\!z\!\in\!\mbR^n\!:\! \abs{z_i}\!\leq\! \check\rho_r^i\right\}\cap \mcX$ and update $b_{f,\mcX_r}$ \label{line:Xr-update-in-l1rg}
\If{$b_{f,\mcX_r}^{old} - b_{f,\mcX_r}>\textup{tol}$}
\State Set $b_{f,\mcX_r}^{old} = b_{f,\mcX_r}$ and go to step \ref{line:Ti}
\EndIf
\State Set $\mcX_a = \{ z\in\mbR^n: \abs{z_i}\leq \rho^i,\ i\in\Zn\}\cap \mcX$ \label{line:Xa-update-in-l1rg}
\While{constraint \cref{eq:T-constraint} does not hold with $\mcC(s)$ from step~\ref{line:Xr-b_f-under-csts} and $\mcX_a$ from step~\ref{line:Xa-update-in-l1rg},}
\State Decrease $T$ 
\EndWhile \Comment{$T$ is updated}
\label{line:T}
\State With $\mcC(s)$, $\mcX_r$, and $T$ from step \ref{line:T}, compute $\gamma_0(T)$ from \cref{eq:gamma_0-T-defn} and $\gamma_2$ from \cref{eq:gamma2-defn}.
\For{$j=1,\dots,m$}
\State Compute $\rho_{u_\at}^j$ and $\tilde \rho_u^j$ according to \cref{eq:tilrho-u-j-defn}
 \EndFor 
\State Compute $\tilde \mcX$ and $\tilde \mcU$ with $\{\tilde \rho^i\}_{i\in\Zn}$ and $\{\tilde \rho_{u}^j\}_{j\in\Zm}$ 
via \cref{eq:xui-xuni-bnd-from-trans-w-tilX-tilU-defn}

\State Set $\mcX_\nt \trieq \mcX \ominus \tilde \mcX$ and $\mcU_\nt \trieq \mcX \ominus \tilde \mcU$
\Ensure{Tightened bounds for the nominal system \cref{eq:nominal system} and an \loneAC~to compensate for uncertainties}
\end{algorithmic}
\end{algorithm}

With the tightened constraints solved from \cref{al1}, we now introduce the design of MPC for the nominal system and how to achieve $u_\text{opt}$. 
At any time $\tau$ with state $x_\nt(\tau)$, $u_\text{opt}(\tau)$ is defined by the solution of the following optimization problem:
\begin{align}\label{eq:mpc}
    &\min_{u(\cdot)}\mathbf{J}(x_\nt(\tau),u(\cdot))  \nonumber\\
    \text{s.t.}~     &\dot x(t)  = A_m x(t) +  Bu(t), x(0)=x_\nt(\tau), t\in[0,T_f],\nonumber\\
    &x(t)\in  \mcX_\nt, t\in[0,T_f], \nonumber \\
    &K_x x(t)+u(t)\in  \mcU_\nt, t\in[0,T_f],
\end{align}
where $T_f$ is the time horizon and given any control input trajectory $u(\cdot)$ over $t\in[0,T_f]$, $\mathbf{J}(x_\nt(\tau),u(\cdot))$ is designed as
\begin{align}
    \mathbf{J}(x_\nt(\tau),u(\cdot))=l_f(x(T_f))+\int_{0}^{T_f} l(x(t),u(t)) \,dt 
\end{align}
with custom terminal cost function $l_f$ and running cost function $l$.
Finally, we can achieve $u_\text{opt}(\tau+t)$ by setting $u_\text{opt}(\tau+t)=u^*(t)$ for $0\leq t < t_\delta$, where $t_\delta$ is a sufficiently small update period and $u^*(\cdot)$ is the solution to the optimization problem \cref{eq:mpc} given $x_\nt(\tau)$. In defining MPC law by the solution of a continuous-time optimal control problem, we follow \cite{magni2004stabilizing}.

\begin{assumption}
We assume that the optimization problem \cref{eq:mpc} is recursively feasible and the states of the controlled nominal system are bounded.
\end{assumption}
\begin{remark}
    Since the optimization problem in \cref{eq:mpc} is a standard MPC problem for a nominal system without any uncertainties, existing techniques, e.g., based on terminal constraints, can be directly applied to ensure/check the recursive feasibility and boundedness of the solution of \cref{eq:mpc} \cite{rawlings2017mpc-book}.  
\end{remark}

\section{Simulation Case Study}
We now verify the performance of the proposed UC-MPC on the case study of controlling the longitudinal motion of an F-16 aircraft adopted from \cite{sobel1985design-pitch}. The model has been slightly simplified by neglecting the actuator dynamics. The open-loop dynamics are given by 
\begin{align}
  \hspace{-1mm}  &\dot x = \begin{bmatrix}
    0 & 0.0067& 1.34 \\
    0 & -0.869 & 43.2 \\
    0 & 0.993 & -1.34
    \end{bmatrix} x \nonumber\\
    &\!+\!\begin{bmatrix}
    0.169 & 0.252 \\
    -17.3& -1.58 \\
    -0.169 & -0.252 
    \end{bmatrix}\! (u\!+\!f(t,x)) 
    \!+\!\begin{bmatrix}
        0.1061 \\0 \\ 0.1061
    \end{bmatrix} \!w(t), \label{eq:ol-dynamics-f16}
\end{align}
where the state $x(t) = [\gamma(t), q(t), \alpha(t)]^\top$ consists of the flight path angle, pitch rate, and angle of attack, the control input $u(t)=[\delta_e(t), \delta_f(t)]$ consists of the elevator deflection and flaperon deflection, and
\begin{align}
    \label{eq:f}
    f(t,x) &= [-1.44\sin(0.4\pi t)-0.18\alpha^2, 0.18-0.36\alpha]^\top, \nonumber\\
    w(t)&=\sin(0.6\pi t)
\end{align}
are assumed uncertainties that depend on both time and $\alpha$. 
The output vector is given by $y(t) = [\theta(t),\gamma(t)]^\top$, where
$\theta(t) = \gamma(t) + \alpha(t)$ is the pitch angle, and we want the output vector $y(t)$ to track the reference trajectory $r(t) = [\theta_c(t),\gamma_c(t)]^\top$, where $\theta_c$ and $\gamma_c$ are the desired pitch
angle and flight path angle, respectively. 
The system is subject to state and control constraints:
\begin{equation}\label{eq:cts-F16}
   \abs{\alpha(t)} \le 4 \!\textup{ deg},\   \abs{\delta_e(t)} \le 25 \!\textup{ deg}, \  \abs{\delta_f(t)} \le 22 \! \textup{ deg}.  
\end{equation}
Furthermore, we assume 
\begin{equation}\label{eq:F16-init}
    \linfnorm{r}\leq 10, \quad x(0)\in \mcX_0 = \Omega(0.1).
\end{equation}
Through simple simulations, we found that given the above constraints and uncertainty formulation, $|\gamma(t)| \le 10$ and $|q(t)|\le 100$. As a result, following the convention in \cref{eq:constraints}, we can write the state constraint as $x(t)\in\mcX\trieq [-10,10]\times [-100,100]\times[-4,4]$. The feedback gain of the baseline controller was set as $K_x = [3.25,0.891,7.12;-6.10,-0.898,-10.0]$, and $A_m=A+BK_x$ is a Hurwitz matrix. 

\subsection{UC-MPC Design}\label{sec:sub-l1rg-f16}
According to the formulation of the uncertainty in \cref{eq:f}, given any set $\mcZ$, $L_{f_1,\mcZ} = 0.36\max_{\alpha\in\mcZ_3}\abs{\alpha} $, $L_{f_2,\mcZ} =0.36$,  $b_{f_1,\mcZ} = 1.44+0.18\max_{\alpha\in\mcZ_3}\alpha^2$, $b_{f_2,\mcZ} =0.18+0.36\max_{\alpha\in\mcZ_3}\abs{\alpha}$ satisfy \cref{assump:lipschitz-bnd-fi}. In addition, $b_w=1$ satisfies \cref{assump:lipschitz-bnd-fi}. For design of the \loneAC~in \cref{eq:l1-control-law,eq:adaptive_law,eq:state-predictor}, we selected $A_e = -10I_3$ and parameterized the filter as $\mcC(s)= \frac{k_f}{s+k_f}I_2$, where the bandwidth for both input channels was set as $k_f=200$. We further set $\gamma_1=0.02$ and the estimation sample time $T$ to be $10^{-7}$ sec, which satisfies \cref{eq:T-constraint}. When applying the scaling technique, we set $\Tau_x^i[k]=0.01$ for each $i,k\in\mathbb{Z}_1^3$ and $k\neq i$, which satisfies \cref{eq:Tx-i-cts}. 
The reference command $r(t)$ was set to be $[9,6.5]$ deg for $t\in[0,7.5]$ sec, and $[0,0]$ deg for $t\in[7.5,15]$ sec. The calculated individual bounds following \cref{al1} are listed as follows: $\tilde{\rho}=[0.08,0.84,0.09]$, $\tilde{\rho}_u=[6.01,3.83]$, and $\rho_{u_a}=[4.34;1.65]$. Thus, the constrained domains for the nominal system were designed as $x(t)\in\mcX_\nt\trieq [-9.92,9.92]\times [-99.17,99.17]\times[-3.91,3.91]$ and $u(t)\in\mcU_\nt\trieq [-18.98,18.98]\times [-18.16,18.16]$. For the design of MPC for the nominal system, the time horizon was selected as 0.2 sec, and the cost function in \cref{eq:mpc} at time $\tau$ was selected as
\begin{align*}
    \mathbf{J}(x_\nt(\tau),u(\cdot))&=\int_{0}^{0.2}100\|Cx(t)-r(\tau+t)\|\\&+ \|u(t)\|+
    100\|\dot u(t)\| \,dt.
\end{align*}
We include $\|\dot u(t)\|$ to penalize control inputs with large time derivatives to achieve a relatively smooth control input trajectory. 

\subsection{Simulation Results}
\begin{figure}[t]
    \centering
    \includegraphics[width=1\columnwidth]{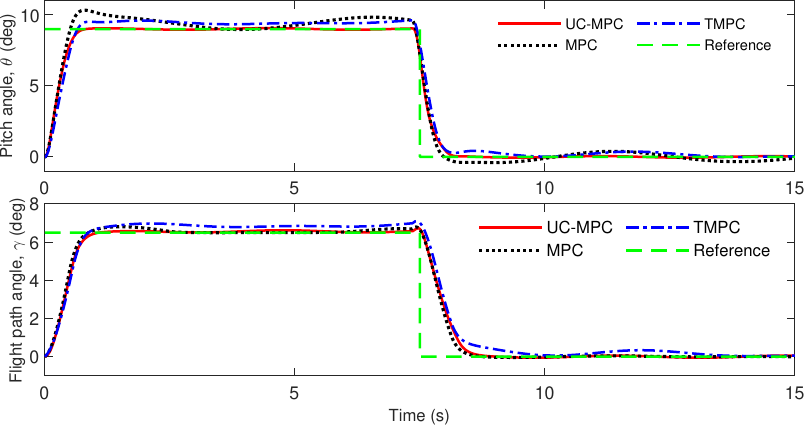} 
    \vspace{-5mm}
    \caption{Tracking performance under MPC, TMPC and UC-MPC (ours).}\label{fig:tracking}
        \vspace{-3mm}
\end{figure}

\begin{figure}[t]
    \centering
    \includegraphics[width=1\columnwidth]{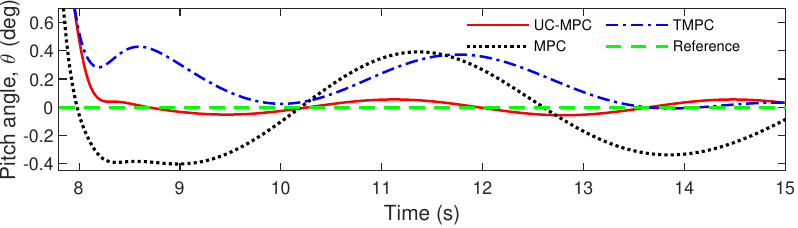} 
    \vspace{-5mm}
    \caption{Zoomed-in view of tracking performance on $\theta$ under MPC, TMPC and UC-MPC.}\label{fig:tracking_zoom}
        \vspace{-3mm}
\end{figure}

\begin{figure}[!t]
    \centering
    \includegraphics[width=1\columnwidth]{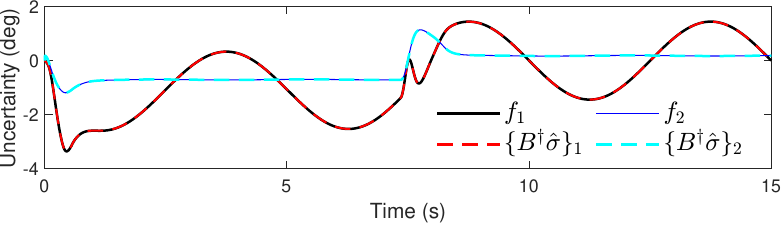} 
        \vspace{-5mm}
    \caption{Actual and estimated uncertainties under UC-MPC. For $i=1,2$, the symbols $f_j$ and $\{B^\dagger\hat\sigma\}_i$ denotes the $i$th element of $f$ and $B^\dagger\hat\sigma$, respectively.}    \label{fig:uncert}
        \vspace{-3mm}
\end{figure}

\begin{figure}[!t]
    \centering
    \includegraphics[width=1\columnwidth]{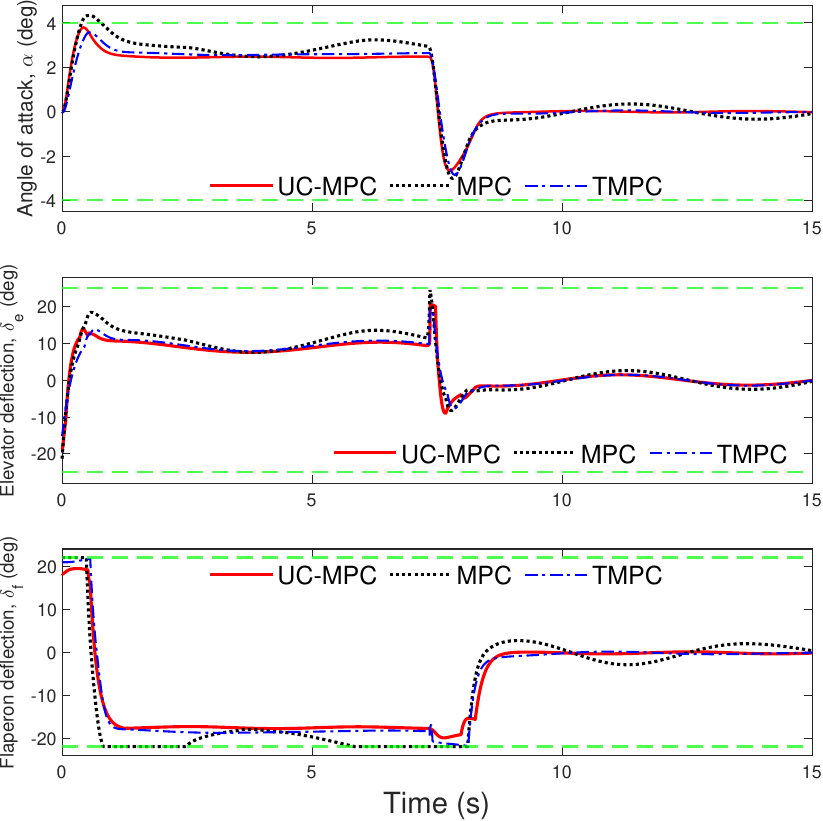}
    \caption{Trajectories of constrained states (top), and control inputs (middle and bottom) under MPC, TMPC, and UC-MPC. Green dash-dotted lines illustrate the constraints specified in \cref{eq:cts-F16}.}    \label{fig:constraints}
        \vspace{-3mm}
\end{figure}

\begin{figure}[t]
    \centering
    \includegraphics[width=1\columnwidth]{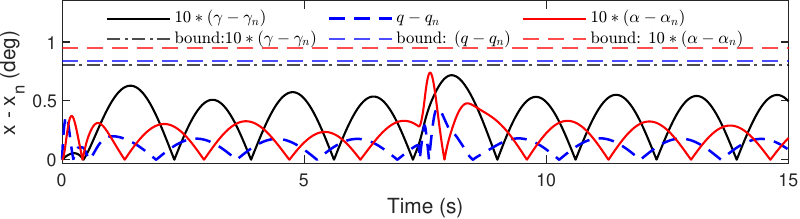} 
        \vspace{-5mm}
    \caption{Trajectories of states of the uncertain system ($x(t)$) under UC-MPC and of the nominal system ($x_\nt(t)$) and their differences. The actual-nominal state errors and bounds for $\gamma(t)$ and $\alpha$ are scaled by 10 for a clear illustration.}    \label{fig:x-xn}
        \vspace{-3mm}
\end{figure}

\begin{figure}[t]
    \centering
    \includegraphics[width=1\columnwidth]{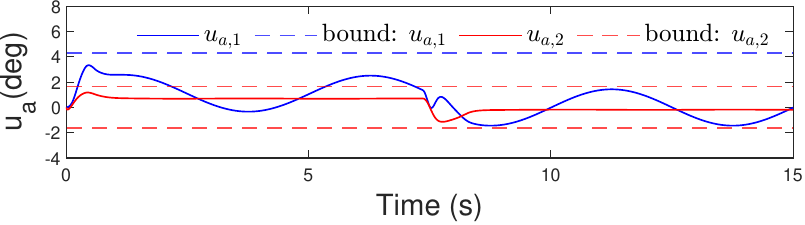} 
        \vspace{-5mm}
    \caption{Adaptive control inputs and theoretical bounds}    \label{fig:ul1}
        \vspace{-3mm}
\end{figure}

\begin{figure}[!t]
    \centering
    \includegraphics[width=1\columnwidth]{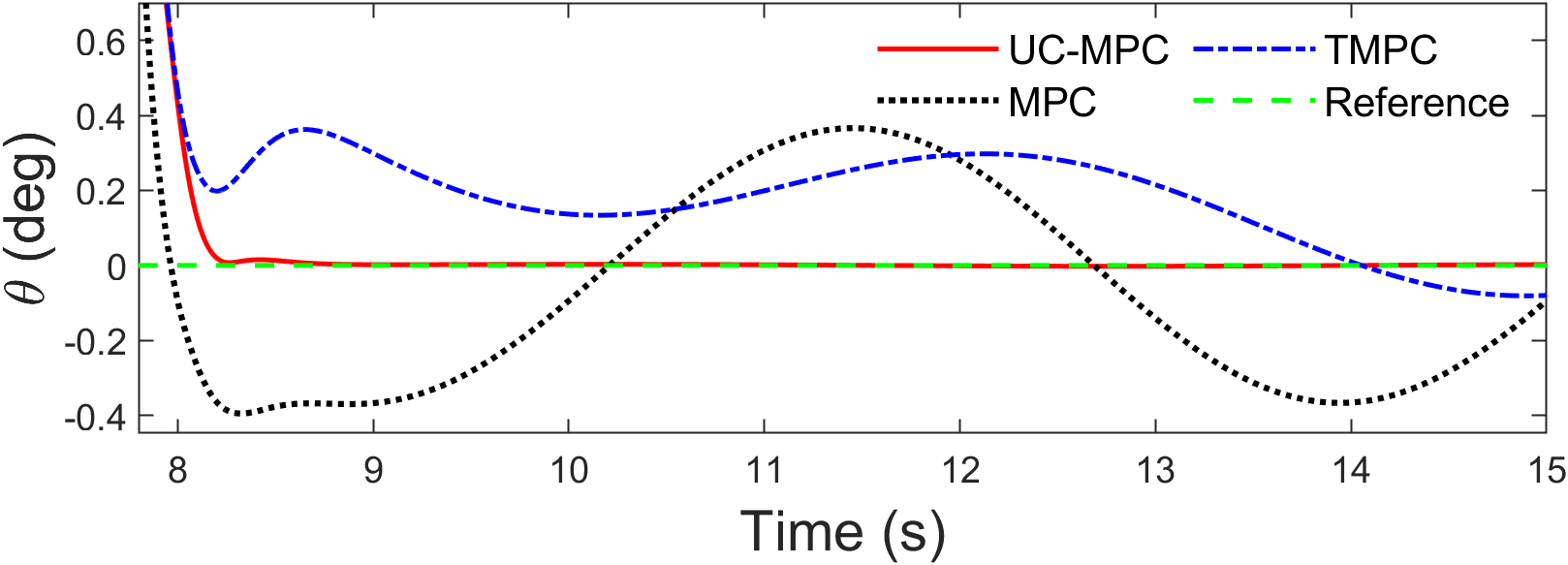} 
    \vspace{-5mm}
    \caption{Zoomed-in view of tracking performance on $\theta$ under MPC, TMPC and UC-MPC when $w(t)=0$}\label{fig:tracking_zoom_no_w}
        \vspace{-3mm}
\end{figure}

For comparison, we also implemented a vanilla MPC, and a Tube MPC (TMPC) \cite{mayne2005robust}. For UC-MPC, the control law follows \cref{eq:total-control-law}: $u(t)=K_xx(t)+u_\text{opt}(t)+u_a(t)$ where $u_\text{opt}(t)$ is achieved from solving the optimization \cref{eq:mpc} with constrained domain $\mcX_\nt$ and $\mcU_\nt$ calculated previously. For implementing the \loneAC, we used a sample time of $10^{-4}$ sec instead of $10^{-7}$ sec used in~\cref{sec:sub-l1rg-f16} for deriving the theoretical bounds. We will show that the theoretical bounds derived under a sample time of $10^{-7}$ sec still hold in simulation, which uses a sample time of $10^{-4}$ sec.
For the vanilla MPC method, the control input $u(t)=u_\text{opt}(t)$ is solved using the nominal system at each time step. The optimization from the vanilla MPC becomes infeasible during the simulation as the nominal system is leveraged to solve for the control input without considering the uncertainty. To avoid infeasibility in completing the simulation, we included the state constraints as soft constraints that would reduce to the original constraints in case they were feasible. For TMPC, we follow the control law $u(t)=u_\text{opt}(t)+K_x(x-x_\nt)$ from \cite{mayne2005robust}, where $u_\text{opt}(t)$ is achieved from solving \cref{eq:mpc} with constrained domains $x\in\mathcal{X}\ominus  Z$ and $u\in\mathcal{X}\ominus K_xZ$, where $Z$ is the disturbance invariance set for the controlled uncertain system $\dotx(t)=(A+BK_x)x(t)+Bf(t,x(t))+B_uw(t)$ w.r.t. the uncertainties $f$ and $w$. A discrete-time formulation was adopted in implementing all MPC.
The simulation results are shown in \cref{fig:tracking,fig:tracking_zoom,fig:tracking_zoom_no_w,fig:constraints,fig:uncert,fig:x-xn,fig:ul1}.


Regarding tracking performance, \cref{fig:tracking,fig:tracking_zoom} show that UC-MPC yielded a better tracking performance compared with both MPC and TMPC, achieving a small error between the actual output $y(t)$ and the reference trajectory $r(t)$. This is because the adaptive control input $u_a(t)$ from UC-MPC approximately cancels the matched uncertainty $f(t,x(t))$, and, according to \cref{fig:uncert}, the matched uncertainty estimation from UC-MPC was accurate. Due to the existence of the unmatched uncertainty $w(t)$, from the zoomed-in view in \cref{fig:tracking_zoom}, we can see that the output trajectories from UC-MPC did not precisely follow the reference trajectories in steady-state, exhibiting an oscillation that has the same frequency as $w(t)$. In contrast, the output trajectories under MPC and TMPC were influenced by both the matched uncertainty and unmatched uncertainty, resulting in a larger tracking error.

Regarding constraints enforcement, 
\cref{fig:constraints} shows that both TMPC and UC-MPC successfully enforced all constraints due to the constraint tightening. In contrast, the output of MPC violated the state constraint as uncertainties were ignored during optimization without constraint tightening. 
It is also worth mentioning that TMPC enforced the constraints at the cost of larger steady-state tracking error, as shown in \cref{fig:tracking_zoom}. 

The trajectories of the actual state $x(t)$ and the nominal state $x_\nt(t)$, as well as the error between them, and the derived individual bounds from \loneAC~are presented in \cref{fig:x-xn}. 
Under UC-MPC, it is evident that the actual states consistently remained in close proximity to the nominal states, exhibiting a difference smaller than the computed individual bound. 
Similarly, according to \cref{fig:ul1}, the adaptive control inputs $u_\at(t)$ remained within the theoretical bounds calculated according to \cref{eq:ua-i-bnd-w-Ua-defn}. 

In addition to the simulation above, we also conducted an experiment {\bf in the presence of only matched uncertainties}, i.e., $w(t)=0$. Under such a scenario, as shown in \cref{fig:tracking_zoom_no_w}, the proposed UC-MPC could successfully compensate for the uncertainty and achieve almost perfect tracking in steady state. Achieving such a high-accuracy tracking performance is not possible with existing robust or tube MPC methods, while the tracking performance of adaptive MPC approaches will also be limited in the presence of inaccurate parameter estimation.



\section{Conclusion}
In this paper, we introduced an uncertainty compensation-based robust adaptive MPC framework, denoted as UC-MPC, for linear systems with both matched and unmatched uncertainties subject to both state and input constraints. Our approach leverages an adaptive controller to actively estimate and compensate for matched uncertainties, ensuring uniform performance bounds on the error between the states and inputs of the actual system and those of an uncertainty-free system. Then, the uniform performance bounds are used to tighten the constraints of the actual system, and an MPC problem is designed for the nominal system with the tightened constraints. Simulation results on a flight control problem demonstrate the efficacy of the proposed UC-MPC in achieving improved performance compared with existing methods while enforcing constraints.


\bibliographystyle{ieeetr}
\bibliography{full}

\begin{thebibliography}{10}

\bibitem{camacho2013mpc-book}
E.~F. Camacho and C.~B. Alba, {\em Model predictive control}.
\newblock Springer Science \& Business Media, 2013.

\bibitem{rawlings2020mpc-book}
J.~B. Rawlings, D.~Q. Mayne, and M.~M. Diehl, {\em {Model Predictive Control:
  Theory, Computation, and Design, 2nd Ed.}}
\newblock Nob Hill Publishing, 2020.

\bibitem{yu2014inherent}
S.~Yu, M.~Reble, H.~Chen, and F.~Allg{\"o}wer, ``Inherent robustness properties
  of quasi-infinite horizon nonlinear model predictive control,'' {\em
  Automatica}, vol.~50, no.~9, pp.~2269--2280, 2014.

\bibitem{kerrigan2001robust-mpc}
E.~C. Kerrigan, {\em Robust constraint satisfaction: {Invariant} sets and
  predictive control}.
\newblock PhD thesis, University of Cambridge, 2001.

\bibitem{langson2004robust-mpc-tube}
W.~Langson, I.~Chryssochoos, S.~Rakovi{\'c}, and D.~Q. Mayne, ``Robust model
  predictive control using tubes,'' {\em Automatica}, vol.~40, no.~1,
  pp.~125--133, 2004.

\bibitem{rakovic2005robust-mpc}
S.~Rakovic, {\em Robust control of constrained discrete time systems:
  {Characterization} and implementation}.
\newblock PhD thesis, University of London, 2005.

\bibitem{mayne2006robust-mpc}
D.~Q. Mayne, S.~V. Rakovi{\'c}, R.~Findeisen, and F.~Allg{\"o}wer, ``Robust
  output feedback model predictive control of constrained linear systems,''
  {\em Automatica}, vol.~42, no.~7, pp.~1217--1222, 2006.

\bibitem{mayne2011tube-mpc-nonlinear}
D.~Q. Mayne, E.~C. Kerrigan, E.~Van~Wyk, and P.~Falugi, ``Tube-based robust
  nonlinear model predictive control,'' {\em International Journal of Robust
  and Nonlinear Control}, vol.~21, no.~11, pp.~1341--1353, 2011.

\bibitem{kohler2020computationally-rmpc}
J.~K{\"o}hler, R.~Soloperto, M.~A. M{\"u}ller, and F.~Allg{\"o}wer, ``A
  computationally efficient robust model predictive control framework for
  uncertain nonlinear systems,'' {\em IEEE Transactions on Automatic Control},
  vol.~66, no.~2, pp.~794--801, 2020.

\bibitem{lopez2019dynamic-tube-mpc}
B.~T. Lopez, J.-J.~E. Slotine, and J.~P. How, ``Dynamic tube {MPC} for
  nonlinear systems,'' in {\em Proceedings of American Control Conference},
  pp.~1655--1662, 2019.

\bibitem{rakovic2018handbook}
S.~V. Rakovic and W.~S. Levine, ``Handbook of model predictive control,'' 2018.

\bibitem{adetola2011robust}
V.~Adetola and M.~Guay, ``Robust adaptive mpc for constrained uncertain
  nonlinear systems,'' {\em International Journal of Adaptive Control and
  Signal Processing}, vol.~25, no.~2, pp.~155--167, 2011.

\bibitem{sasfi2023robust}
A.~Sasfi, M.~N. Zeilinger, and J.~K{\"o}hler, ``Robust adaptive mpc using
  control contraction metrics,'' {\em Automatica}, vol.~155, p.~111169, 2023.

\bibitem{lorenzen2017adaptive}
M.~Lorenzen, F.~Allg{\"o}wer, and M.~Cannon, ``Adaptive model predictive
  control with robust constraint satisfaction,'' {\em IFAC-PapersOnLine},
  vol.~50, no.~1, pp.~3313--3318, 2017.

\bibitem{zhang2020adaptive}
K.~Zhang and Y.~Shi, ``Adaptive model predictive control for a class of
  constrained linear systems with parametric uncertainties,'' {\em Automatica},
  vol.~117, p.~108974, 2020.

\bibitem{zhao2023integrated}
P.~Zhao, I.~Kolmanovsky, and N.~Hovakimyan, ``Integrated adaptive control and
  reference governors for constrained systems with state-dependent
  uncertainties,'' {\em IEEE Transactions on Automatic Control}, 2023.

\bibitem{hovakimyan2010L1-book}
N.~Hovakimyan and C.~Cao, {\em {$\mathcal{L}_1$ Adaptive Control Theory:
  Guaranteed Robustness with Fast Adaptation}}.
\newblock Philadelphia, PA: Society for Industrial and Applied Mathematics,
  2010.

\bibitem{pereida2021robust}
K.~Pereida, L.~Brunke, and A.~P. Schoellig, ``Robust adaptive model predictive
  control for guaranteed fast and accurate stabilization in the presence of
  model errors,'' {\em International Journal of Robust and Nonlinear Control},
  vol.~31, no.~18, pp.~8750--8784, 2021.

\bibitem{magni2004stabilizing}
L.~Magni and R.~Scattolini, ``Stabilizing model predictive control of nonlinear
  continuous time systems,'' {\em Annual Reviews in Control}, vol.~28, no.~1,
  pp.~1--11, 2004.

\bibitem{rawlings2017mpc-book}
J.~B. Rawlings, D.~Q. Mayne, and M.~Diehl, {\em Model Predictive Control:
  Theory, Computation, and Design}, vol.~2.
\newblock Madison, WI: Nob Hill Publishing, 2017.

\bibitem{sobel1985design-pitch}
K.~M. Sobel and E.~Y. Shapiro, ``A design methodology for pitch pointing flight
  control systems,'' {\em Journal of Guidance, Control, and Dynamics}, vol.~8,
  no.~2, pp.~181--187, 1985.

\bibitem{mayne2005robust}
D.~Q. Mayne, M.~M. Seron, and S.~Rakovi{\'c}, ``Robust model predictive control
  of constrained linear systems with bounded disturbances,'' {\em Automatica},
  vol.~41, no.~2, pp.~219--224, 2005.

\end{thebibliography}

\appendix
Before presenting the proofs of the lemmas presented in this paper, we first introduce the following lemmas.
\begin{lemma}\label{lem:L1-Linf-relation}
\cite[Lemma 2]{zhao2023integrated}
For a stable proper MIMO system $\mathcal H(s)$ with states $x(t)\in \mbR^n $, inputs $u(t)\in \mbR^m$ and outputs $y(t)\in \mbR^p$, under zero initial states, i.e., $x(0) =0$, we have 
$\linfnormtruc{y}{\tau}\leq \lonenorm{\mathcal H(s)}\linfnormtruc{u}{\tau}$, for any $\tau\geq 0$. Furthermore, for any matrix $\Tau\in \mbR^{q\times p}$,  we have $\linfnorm{\Tau \mathcal H(s)}\leq \infnorm{\Tau}\linfnorm{\mathcal H(s)}$. 
\end{lemma}

\begin{lemma} \label{lem:ref-xr-ur-bnd}
For the closed-loop reference system in \cref{eq:ref-system} subject to \cref{assump:lipschitz-bnd-fi} and the stability condition in  \cref{eq:l1-stability-condition}, we have
\vspace{-4mm}\\
\begin{align}
 \linfnorm{x_\textup{r}} &<\rho_r \label{eq:xref-bnd}, \\ \linfnorm{u_{\textup{r}}} &<\rho_{ur}, \label{eq:uref-bnd}
\end{align}where $\rho_r$ is introduced in \cref{eq:l1-stability-condition}, and $\rho_{ur}$ is defined in \cref{eq:rho_ur-defn}.
\end{lemma}
\begin{proof}
For notation brevity, we define:
\begin{equation}\label{eq:eta-eta_r-defn}
    \eta(t) \trieq f(t,x(t)),\quad \eta_\rt(t) \trieq f(t,x_\rt(t)).
\end{equation}
Let's first rewrite the dynamics of the reference system in \cref{eq:ref-system} in the Laplace domain:
\begin{align}\label{eq:xr-expression-w-G-f}
   \xr(s) = \mcG_{xm}(s)\laplace{f(t,x_\textup{r}(t))} \nonumber + \mcH_{xm}(s)u_\text{opt}(s)\\+ \mcH_{xu}(s)w(s) + \xin(s).
\end{align}
Considering \cref{eq:rho_in}, $A_m$ is Hurwitz and $\mcX_0$ is compact, we have $\linfnorm{\xin}\leq \rhoin$ according to \cref{lem:L1-Linf-relation}.
As a result, according to \cref{lem:L1-Linf-relation} and because $\linfnorm{w}\leq b_w$, for any $\xi>0$, we have 
\begin{align}\label{eq:x_r_linfnorm_truc_bound}
    \linfnormtruc{x_\textup{r}}{\xi}\! &\leq \lonenorm{\mcG_{xm}(s)} \!\linfnormtruc{\eta_\rt}{\xi}\! +\! \lonenorm{\mcH_{xm}(s)}\linfnorm{u_\text{opt}}\nonumber\\&+ \! \lonenorm{\mcH_{xu}(s)}\!b_w\!+\!\linfnorm{\xin},
\end{align}
where $\eta_\rt(t)$ is defined in \cref{eq:eta-eta_r-defn}. Assume by contradiction that \cref{eq:xref-bnd} is not true. Since $x_\textup{r}(t)$ is continuous and $\infnorm{x_\textup{r}(0)}<\rho_r$, there exists a $\tau\!>\!0$ such that 
\begin{equation}
    \infnorm{x_\textup{r}(t)}<\rho_r, \ \forall t\in[0,\tau),\ \textup{and}\ \infnorm{x_\textup{r}(\tau)}=\rho_r,
\end{equation} 
which implies 
$x_\rt(t)\in \Omega(\rho_r)$ for any $t$ in $[0,\tau]$. With \cref{eq:bnd-f} from \cref{assump:lipschitz-bnd-fi}, it follows that
\begin{equation}\label{eq:f-xr-rhor-tau}
    \linfnormtruc{\eta_\rt}{\tau}\leq b_{f,\Omega(\rho_r)}.
\end{equation}
By plugging the inequality above into  \cref{eq:x_r_linfnorm_truc_bound}, we have 
\begin{align}
    \rho_r &\leq \lonenorm{\mcG_{xm}(s)} b_{f,\Omega(\rho_r)}+ \lonenorm{\mcH_{xm}(s)}\linfnorm{u_\text{opt}} \nonumber\\&+ \lonenorm{\mcH_{xu}(s)}b_w +  \rho_\textup{in},
\end{align}
which contradicts the condition \cref{eq:l1-stability-condition}. Therefore, \cref{eq:xref-bnd} is true. Equation \cref{eq:uref-bnd} immediately follows from \cref{eq:xref-bnd} and \cref{eq:ref-system}. 
\end{proof}

\begin{lemma}\label{lem:xtilde-bnd}
Given the uncertain system \cref{eq:dynamics-uncertain} subject to \cref{assump:lipschitz-bnd-fi}, the state predictor  \cref{eq:state-predictor} and the adaptive law \cref{eq:adaptive_law}, if 
\begin{equation}\label{eq:x-u-tau-bnd-assump-in-lemma}
    \linfnormtruc{x}{\tau}\leq \rho, \quad \linfnormtruc{u_\at}{\tau}\leq \rho_{u_\at},
\end{equation}
with $\rho$ and $\rho_{u_\at}$ defined in \cref{eq:rho-u-defn,eq:rho-defn}, respectively, then 
\begin{align}
  \linfnormtruc{\tilx}{\tau} \leq \gamma_0(T). \label{eq:tilx_tau-leq-gamma0}
\end{align}
\end{lemma}

\begin{proof}
Based on \cref{eq:x-u-tau-bnd-assump-in-lemma}, we have $x(t)\in \Omega(\rho)$ for any $t$ in $[0,\tau]$. Due to \cref{eq:bnd-f} from \cref{assump:lipschitz-bnd-fi}, it follows that
\begin{equation}\label{eq:f-bnd-in-0-tau}
    \infnorm{f(t,x(t))} =\infnorm{\eta(t)} \leq b_{f,\Omega(\rho)},\quad \forall t\in [0,\tau].
\end{equation} 
From \cref{eq:dynamics-uncertain,eq:state-predictor}, the prediction error dynamics are given by
{\begin{equation}\label{eq:prediction-error}
\begin{aligned}
      \dot{\tilx}(t) &= A_e \tilx(t) + \hsigma(t) - B f(t,x(t))-\Bu w(t). 
\end{aligned}
\end{equation}}
For any $0\leq t <T$ and $i\in\mbZ_0$, due to \cref{eq:prediction-error}, we have
{
\begin{align}
    \tilx(iT+t) &= ~e^{A_et}\tilx(iT)+\int_{iT}^{iT+t} e^{A_e(iT+t-\xi)}
\hsigma(iT)   
   d\xi   \nonumber\\&- \int_{iT}^{iT+t} e^{A_e(iT+t-\xi)}\left(B \eta(\xi) + \Bu w(\xi)  \right) d\xi \nonumber \\ 
    & =  ~e^{A_et}\tilx(iT)+\int_{0}^{t} e^{A_e(t-\xi)}
   \hsigma(iT) 
   d\xi   \nonumber \\&- \int_{0}^{t} e^{A_e(t-\xi)}\left(B \eta(iT+\xi) + \Bu w(iT+\xi) \right ) d\xi. 
    \label{eq:tilx-iTplust}
\end{align}}
According to the adaptive law \cref{eq:adaptive_law}, the preceding equality implies
\begin{align}
 &\tilx((i+1)T) \!=\nonumber \\ \!&- \!\int_{0}^{T}\! e^{A_e(T-\xi)}\left(B \eta(iT+\xi) + \Bu w(iT+\xi) \right ) d\xi.
\end{align}
Therefore, considering \cref{eq:f-bnd-in-0-tau}, for any $i\in \mbZ_0$ with $(i+1)T\leq \tau$, we have
\begin{align}
    \infnorm{\tilx((i+1)T)} &\leq \bar \alpha_0(T) b_{f,\Omega(\rho)} +\bar \alpha_1(T) b_w,
\end{align}
where $\bar \alpha_0(T)$ and $\bar \alpha_1(T)$ are defined in \cref{eq:alpha_0-defn} and \cref{eq:alpha_1-defn}, respectively. 
Since $\tilx(0)=0$, 
\begin{align}\label{eq:tilx-iT-bnd}
    \infnorm{\tilx(iT)}\leq \bar \alpha_0(T) b_{f,\Omega(\rho)} +\bar \alpha_1(T) b_w \leq \gamma_0(T), \nonumber \\\; \forall  iT\leq \tau, i\in \mbZ_0.
\end{align}
Now we consider any $t\in(0,T]$ such that $iT+t\leq \tau$ with $i\in \mbZ_0$. From \cref{eq:tilx-iTplust} and the adaptive law \cref{eq:adaptive_law}, we have
\begin{align}
    &\infnorm{\tilx(iT+t)} \leq  \infnorm{e^{A_et}}\infnorm{\tilx(iT)} \hspace{-2cm}\nonumber\\ & +  \int_{0}^{t} \infnorm{e^{A_e(t-\xi)}\Phi^{-1}(T)e^{A_eT}} \infnorm{\tilx(iT)}  d\xi   \nonumber \\
  & +\int_0^t \infnorm{e^{A_e(t-\xi)}B} \infnorm{\eta(iT+\xi)} d\xi \nonumber\\&+ \int_0^t \infnorm{e^{A_e(t-\xi)}\Bu} \infnorm{w(iT+\xi)} d\xi \nonumber \\
 &\leq   \left(\bar \alpha_2(T)+\bar \alpha_3(T)+1 \right) (\bar\alpha_0(T) b_{f,\Omega(\rho)} +\bar \alpha_1(T)b_w) \nonumber\\ &= \gamma_0(T),  \label{eq:tilx-iT+t-bnd} 
\end{align}
where $\bar\alpha_i(T)$ ($i=0,1,2,3$) are defined in \cref{eq:alpha_0-defn,eq:alpha_1-defn,eq:alpha_2-defn}, and the last inequality is partially due to the fact that $\int_0^t \infnorm{e^{A_e(t-\xi)}B}d\xi\leq\int_0^T \infnorm{e^{A_e(T-\xi)}B}\!d\xi\!=\!\bar\alpha_0(T) $ and $\int_0^t \infnorm{e^{A_e(t-\xi)}B_u}d\xi\leq\int_0^T \infnorm{e^{A_e(T-\xi)}B_u}\!d\xi\!=\!\bar\alpha_1(T) $. Equations \cref{eq:tilx-iT-bnd,eq:tilx-iT+t-bnd} imply \cref{eq:tilx_tau-leq-gamma0}. 
\end{proof}

\section{Proofs}
\subsection{Proof of \cref{them:x-xref-bnd}}
\begin{proof}
\cref{eq:xref-x-bnd,eq:uref-u-bnd} are proved by contradiction. Assume that \cref{eq:xref-x-bnd} or \cref{eq:uref-u-bnd} do not hold. Since $\infnorm{x_\textup{r}(0)-x(0)}=0<\gamma_1$, $\infnorm{u_{\textup{r}}(0)-u_\at(0)}=0<\gamma_2$, and $x(t)$, $u_\at(t)$, $x_\textup{r}(t)$ and $u_{\textup{r}}(t)$ are all continuous, there must exist a time instant $\tau$ such that
\begin{equation}
 \hspace{-2mm}   \infnorm{x_\textup{r}(\tau)-x(\tau)} = \gamma_1 \textup{ or }  \infnorm{u_{\textup{r}}(\tau)-u(\tau)} = \gamma_2,
\end{equation}
and
\begin{equation} 
   \hspace{-3mm} \infnorm{x_\textup{r}(t)\!-\!x(t)}\! <\! \gamma_1, \  \infnorm{u_{\textup{r}}(t)\!-\!u(t)} \!<\! \gamma_2, \ \forall t\in [0,\tau).
\end{equation}
It follows that at least one of the following equalities must hold:
\begin{equation}\label{eq:xr-x-ur-u-linfnorm-tau}
    \linfnormtruc{x_\textup{r}-x}{\tau} = \gamma_1, \quad \linfnormtruc{u_{\textup{r}}-u_\at}{\tau} = \gamma_2.
\end{equation}
According to \cref{lem:ref-xr-ur-bnd}, we have $\linfnorm{x_\textup{r}} \leq \rho_r<\rho $ and according to \cref{eq:xr-x-ur-u-linfnorm-tau}, we have $\linfnorm{x}\leq \rho_r+\gamma_1 = \rho$. Further considering \cref{eq:lipschitz-cond-f} that results from \cref{assump:lipschitz-bnd-fi}, we achieve
\begin{equation}\label{eq:f-xr-x-tau-bnd}
 \hspace{-1mm}   \infnorm{f(t,x_\textup{r}(t))\!-\!f(t,x(t))} \!\leq\! L_{f,\Omega(\rho)}\! \linfnormtruc{x_\textup{r}\!-\!x}{\tau}\!, \ \forall t \!\in \![0,\tau].
\end{equation}
The control laws in \cref{eq:l1-control-law} and  \cref{eq:ref-system} indicate that
{
\begin{align}
&u_{\textup{r}}(s) - u_\at(s)  = -\mcC(s)\laplace{f(t,x_\textup{r})-B^\dagger \hsigma(t)} = \nonumber\\& \mcC(s)\laplace{f(t,x)\!-\!f(t,x_\textup{r})} + \mcC(s)(B^\dagger \hsigma(s) \!-\! \laplace{f(t,x)}).\label{eq:omega-ur-u}
\end{align}
From equation \cref{eq:prediction-error}, we have
\begin{equation}\label{eq:hsigma-sigma-s}
    B^\dagger \hsigma(s) - \laplace{f(t,x)}) = B^\dagger(sI_n-A_e)\tilx(s).
\end{equation}}
Considering \cref{eq:dynamics-uncertain}, \cref{eq:l1-control-law} and \cref{eq:hsigma-sigma-s}, we have 
\begin{align}
x(s) \!&= \mcG_{xm}(s) \laplace{f(t,x)} \!+ \!\mcH_{xm}(s)u_\text{opt}(s) \!+\! \mcH_{xu}(s)w(s) \nonumber \\ &+\xin(s) 
     - \mcH_{xm}(s)\mcC(s)B^\dagger(sI_n-A_e)\tilx(s),
\end{align}
  which, together with \cref{eq:xr-expression-w-G-f}, implies
  \begin{align}
x_\textup{r}(s) - x(s) = \mcG_{xm}(s) \laplace{f(t,x_\textup{r})-f(t,x)} \nonumber \\
     +\mcH_{xm}(s)\mcC(s)B^\dagger(sI_n-A_e)\tilx(s).
\end{align}
Therefore, due to \cref{eq:f-xr-x-tau-bnd} and \cref{lem:xtilde-bnd}, we have 
\begin{align}
    \linfnormtruc{x_\textup{r}-x}{\tau} &\leq  \lonenorm{\mcG_{xm}} L_{f,\Omega(\rho)} \linfnormtruc{x_\textup{r}-x}{\tau}  \\&+\! \lonenorm{\mcH_{xm}(s)\mcC(s)B^\dagger(sI_n-A_e)}\!\gamma_0(T).\nonumber
\end{align}
The preceding equation, together with \cref{eq:l1-stability-condition-Lf}, leads to 
\begin{align}
   \hspace{-2mm} \linfnormtruc{x_\textup{r}-x}{\tau} &\!\leq\! \frac{\lonenorm{\mcH_{xm}(s)\mcC(s)B^\dagger(sI_n\!-\!A_e)}}{1-    \lonenorm{\mcG_{xm}}  L_{f,\Omega(\rho)}} \gamma_0(T),
\end{align}
which, together with the sample time constraint \cref{eq:T-constraint}, indicates that 
\begin{equation}\label{eq:xr-x<gamma1}
    \linfnormtruc{x_\textup{r}-x}{\tau}  < \gamma_1. 
\end{equation}

On the other hand, it follows from \cref{eq:f-xr-x-tau-bnd,eq:omega-ur-u,eq:hsigma-sigma-s,eq:xr-x<gamma1} that
\begin{align*}
  &\linfnormtruc{ u_{\textup{r}}- u_\at}{\tau}   \leq  \lonenorm{\mcC(s)} L_{f,\Omega(\rho)}  \linfnormtruc{x_\textup{r}-x}{\tau} \\&+ \lonenorm{\mcC(s)B^\dagger(sI_n-A_e)}\linfnormtruc{\tilx}{\tau} \nonumber \\
    & < \lonenorm{\mcC(s)} L_{f,\Omega(\rho)} \gamma_1 + \lonenorm{\mcC(s)B^\dagger(sI_n-A_e)}\gamma_0(T).
\end{align*}
Further considering the definition in \cref{eq:gamma2-defn}, we have
\begin{equation}\label{eq:ur-u<gamma2}
    \linfnormtruc{ u_{\textup{r}}- u_\at}{\tau}<\gamma_2.
\end{equation}
Now, \cref{eq:xr-x<gamma1} and \cref{eq:ur-u<gamma2} contradict the \cref{eq:xr-x-ur-u-linfnorm-tau}, which shows that \cref{eq:xref-x-bnd,eq:uref-u-bnd} holds. The bounds in \cref{eq:x-bnd,eq:ua-bnd} follow directly from  \cref{eq:xref-x-bnd,eq:uref-u-bnd,eq:xref-bnd,eq:uref-bnd} and the definitions of $\rho$ and $\rho_{u_\at}$ in \cref{eq:rho-defn,eq:rho-u-defn}.
The proof is complete. 
\end{proof}

\subsection{Proof of \cref{lem:ref-id-bnd}}
\begin{proof}
Considering \cref{eq:nominal system,eq:ref-system}, we have 
\begin{align}\label{eq:xr-xn-expression}
\hspace{-2mm}   \xr(s)-\xn(s) &= G_{xm}(s) \laplace{f(t,x_\textup{r})} + \mcH_{xu}(s)w(s) \nonumber\\&= G_{xm}(s) \laplace{\eta_\textup{r}(t)}+ \mcH_{xu}(s)w(s).
\end{align}
From \cref{lem:ref-xr-ur-bnd}, we have $x_\rt(t) \in \Omega(\rho_r)$ for any $t\geq0$. Further because
 \cref{eq:bnd-f} that results from \cref{assump:lipschitz-bnd-fi}, it follows that $\linfnorm{\eta_\rt}\leq b_{f,\Omega(\rho_r)} $, which, together with \cref{eq:xr-xn-expression}, leads to \cref{eq:xref-xid-bnd}. 
\end{proof}

\subsection{Proof of \cref{lem:refine_bnd_xi_w_Txi}}
\begin{proof}
Given any $\Tau_x^i$ satisfying \cref{eq:Tx-i-cts} with an arbitrary $i\in\Zn$, it follows that $\infnorm{\Tau_x^i}= 1$. As a result, using the transformation \cref{eq:coordinate-trans} and considering \cref{eq:Hxm-Hxv-Gxm-check-defn,eq:check-rhoin-defn} and \cref{lem:L1-Linf-relation}, the following inequalities hold
\begin{subequations}\label{eq:Hxm-Hxv-Gxm-check-original-relation}
\begin{align}
      \hspace{-4mm}   \linfnorm{\mcH_{\check xm}^i(s)}\! & \! \leq \! \infnorm{\Tau_x^i}\! \linfnorm{\mcH_{xm}(s)}\! \!=\! \linfnorm{\mcH_{xm}(s)}\!,  \label{eq:Hxm-check-original-relation} \\
   \hspace{-4mm}   \linfnorm{\mcH_{\check xu}^i(s)}\! & \! \leq \! \infnorm{\Tau_x^i} \! \linfnorm{\mcH_{xu}(s)} \!=\!  \linfnorm{\mcH_{xu}(s)}\!, \label{eq:Hxv-check-original-relation}\\
\hspace{-2mm}  \linfnorm{\mcG_{\check xm}^i(s)}\! & \! \leq \!  \infnorm{\Tau_x^i}  \!\linfnorm{\mcG_{xm}(s)} \!=\!  \linfnorm{\mcG_{xm}(s)}\!, \label{eq:Gxm-check-original-relation} \\
\hspace{-2mm}   \check \rho_\textup{in}^i & \! \leq \! \infnorm{\Tau_x^i}  \rhoin \!=\! \rhoin. \label{eq:rhoin-check-original-relation}
\end{align}
\end{subequations}
The property of $\xr(t)\in \mcX_r$ for any $t\geq0$ from \cref{lem:ref-xr-ur-bnd} and \cref{eq:coordinate-trans} together imply $\check x_\rt(t)\in \check \mcX_r$ for any $t\geq0$, where $\check \mcX_r$ is defined via \cref{eq:check-Z-defn}. 
Considering \cref{eq:f-checkf-relation,eq:check-Z-defn}, for any compact set $\mcX_r$, we have
\begin{equation}\label{eq:b-checkf-f-equal}
    b_{\check f, \check \mcX_r} = b_{f, \mcX_r}.
\end{equation}
Suppose that constants $\rho_r$ and $\linfnorm{u_\text{opt}}$ satisfy \cref{eq:l1-stability-condition}.  
According to \cref{eq:b-checkf-f-equal,eq:Hxm-Hxv-Gxm-check-original-relation}, with $\check \rho_r^i = \rho_r$ and the same $\linfnorm{u_\text{opt}}$, \cref{eq:l1-stability-condition-transformed} is satisfied. 

In addition, if \cref{eq:l1-stability-condition-transformed} holds, through the application of \cref{lem:ref-xr-ur-bnd} to the transformed reference system \cref{eq:ref-system-transformed}, we obtain $\linfnorm{\check x_\rt}\leq \check \rho_r^i$, which further implies that $\abs{\check x_{\rt,i}(t)}\leq \check \rho_r^i$ for any $t\geq 0$. Since $\check x_{\rt,i}(t) =x_{\rt,i}(t) $ due to the constraint \cref{eq:Tx-i-cts} on $\Tau_x^i$, we have \cref{eq:xr-i-bnd-from-trans}.
Equation \cref{eq:xr-i-bnd-from-trans} is equivalent to $\xr(t)\in\mcX_r$ for any $t\geq 0$, with the re-definition of $\mcX_r$ in \cref{eq:Xr-defn}. 
Following the proof of \cref{lem:ref-id-bnd}, we are able to obtain $   \left\| {\check x_\textup{r} - \check x_\nt} \right\|_{{\mathcal L}{_\infty }}  \leq  \lonenorm{\mcG_{ \check xm}} b_{\check f, \check \mcX_r} + \lonenorm{\mcH_{ \check xu}} b_w= \lonenorm{\mcG_{ \check xm} (s)}b_{f, \mcX_r} + \lonenorm{\mcH_{ \check xu}} b_w$, where the equality comes from \cref{eq:b-checkf-f-equal}. Further considering $\check x_{\rt,i}(t) =x_{\rt,i}(t) $ and $\check x_{\nt,i}(t) = x_{\nt,i}(t)$ due to the constraint \cref{eq:Tx-i-cts} on $\Tau_x^i$, we finally have \cref{eq:xri-xni-bnd-from-trans}.
\end{proof}

\subsection{Proof of \cref{them:xi-uai-bnd}}
\begin{proof}
For each $i\in \Zn$, \cref{lem:refine_bnd_xi_w_Txi} implies that 
$\abs{x_{\rt,i}(t)} \leq \check \rho_r^i$ and $\abs{x_\textup{r,i}(t)-x_{\nt,i}(t)} \leq \lonenorm{\mcG_{ \check xm}^i(s)}b_{f,\mcX_r} + \lonenorm{\mcH^i_{ \check xu}} b_w$ for all  $t\geq0$. From \cref{them:x-xref-bnd}, it follows that $\abs{x_{\rt,i}(t) - x_{i}(t) } \leq \gamma_1$ for any $t\geq 0$ and any $i\in \Zn$. Thus, \cref{eq:xi-xni-bnd-from-trans-w-tilX-defn} is true. 
On the other hand, \cref{them:x-xref-bnd} indicates that $\abs{u_{\rt,j}(t) - u_{\at,j}(t) } \leq \gamma_2$ for any $t\geq 0$. Property \cref{eq:ref-system} and the structure of $\mcC(s)$ in \cref{eq:filter-defn} lead to
\begin{equation}
    u_{\rt,j}(s)=-\mcC_j(s) \laplace{f_j(t,\xr(t))}, \quad \forall j\in\Zm.
\end{equation}
Therefore, given a set $\mcX_r$ such that $\xr(t)\!\in\! \mcX_r$ for any $t\!\geq\! 0$, from \cref{assump:lipschitz-bnd-fi,lem:L1-Linf-relation}, the following property holds 
\begin{equation}\label{eq:uar-i-bnd}
 \abs{u_{\rt,j}(t)} \leq \lonenorm{\mcC_j(s)} b_{f_j, \mcX_r}, \quad \forall t\geq 0, \ \forall j\in\Zm.
\end{equation}
Thus, for any $j\in\Zm$, we have \cref{eq:ua-i-bnd-w-Ua-defn}. Finally, since $u(t) - u_\nt(t) = K_x(x(t)-\xn(t)) + u_\at(t)$,
we achieve \cref{eq:uj-unj-bnd-w-tilU-defn}. The proof is complete. 
\end{proof}

\end{document}